\documentclass[aps,prb,a4paper,amssymb,amsmath,reprint,superscriptaddress]{revtex4-2}
\usepackage{amsmath,amssymb}
\usepackage{bm,braket,latexsym,multirow}
\usepackage{bbding,pifont,wasysym}
\usepackage{colortbl,array}
\usepackage{graphicx} 
\usepackage{xcolor}

\usepackage[
    pdftex,
draft=false,
bookmarks=true,
bookmarksnumbered=false,
bookmarksopen=false,
bookmarksopenlevel=4,
colorlinks=true,
anchorcolor=black,
citecolor=blue,
filecolor=black,
linkcolor=black,
linkbordercolor={1 0 0},
urlcolor=violet,
pdfborder={0 0 1},
pdftitle={},
pdfauthor={},
pdfsubject={},
pdfkeywords={},
pdfpagemode=UseThumbs,
]{hyperref}

\usepackage{xspace}
\newcommand{\Pa}{$\mathcal{P}$\xspace}
\newcommand{\T}{$\mathcal{T}$\xspace}
\newcommand{\PT}{$\mathcal{PT}$\xspace}
\newcommand{\sro}{Sr$_2$IrO$_4$\xspace}

\newcommand{\mr}[2]{#1 \, {\mathrm{ \,  #2 \,}}}

\newcommand{\bk}{\bm{k}}

\newcommand{\ud}{\mathfrak{D}}
\newcommand{\bce}{\mathcal{A}}

\begin{document}

\title{Photocurrent response in parity-time symmetric current-ordered states}
\author{Hikaru Watanabe}
\affiliation{RIKEN Center for Emergent Matter Science (CEMS), Wako, Saitama 351-0198, Japan}
\affiliation{Department of Physics, Graduate School of Science, Kyoto University, Kyoto 606-8502, Japan}
\author{Youichi Yanase}
\affiliation{Department of Physics, Graduate School of Science, Kyoto University, Kyoto 606-8502, Japan}
\affiliation{Institute for Molecular Science, Okazaki,444-8585, Japan}
\date{\today}

\begin{abstract}
There is growing interest in the photo-induced generation of rectified current, namely photocurrent phenomenon. While the response was attributed to noncentrosymmetric structures of crystals, the parity violation accompanied by the magnetic ordering, that is, magnetic parity violation, is recently attracting attention as a platform for a photocurrent generator. In this paper, we investigate the photocurrent response in the current-ordered phase, realizing the magnetic parity violation without the spin degree of freedom, although prior studies focused on the parity-violating spin structure. The loop-current order breaks the inversion symmetry while preserving the parity-time-reversal symmetry. With a model of \sro{}, we demonstrate the linearly and circularly polarized light-induced photocurrent responses in the current-ordered state.
Each photocurrent has a distinct tolerance of the scattering rate according to the mechanism for the photocurrent creation. The predicted photocurrent response is comparable to that in prototypical semiconductors. We propose a probe to detect the hidden-ordered phase in \sro{} by the photocurrent response.

\end{abstract}
\maketitle

\section{Introduction}

The photocurrent response (photogalvanic effect), that is, rectified current generation from irradiating lights, has a potential application uplifting our daily life such as photo-sensing and solar panel technologies. Historically, the photocurrent response has been attributed to the internal electric field originating from the ferroelectric order and heterojunction of semiconductors and to the external electric field. On the other hand, a lot of attention has also been paid to the bulk photocurrent response. Apart from a prototypical mechanism for the photocurrent generation, the bulk photocurrent arises from the homogeneous and noncentrosymmetric structure in bulk crystals and shows the characteristic properties of symmetry and frequency dependencies~\cite{Sturman1992Book}. The bulk photocurrent was discovered in a well-known ferroelectric material, BaTiO$_3$~\cite{Koch1976BTO}. Moreover, recent studies have clarified that topological materials may intensely improve the performance of the photocurrent generation~\cite{Liu2020SemimetalReview}. From the theoretical viewpoint, the photocurrent response is formulated by the established perturbative calculations and understood in terms of the quantum geometry of the electronic band structures~\cite{Sipe2000secondorder,Ventura2017,Passos2018,Parker2019,Joao2020-mk,Michishita2020}. The obtained formulas are implemented in first-principle calculations and have successfully explained the photocurrent responses in ferroelectric perovskite compounds~\cite{YoungRappe2012_FirstPrincipleBTO,YoungRappe2012BiFeO3}. 

An essential ingredient for the photocurrent response, the violation of the parity symmetry (\Pa symmetry), is usually provided by the noncentrosymmetric crystal structure. A lot of prior studies have therefore worked on noncentrosymmetric and non-magnetic materials preserving the time-reversal symmetry (\T symmetry) such as ferroelectric compounds and non-magnetic Weyl semimetals. On the other hand, it has recently been shown that the parity violation due to the magnetic order, namely magnetic parity violation, also gives rise to the photocurrent response~\cite{Zhang2019switchable,watanabe2020chiral,Ahn2020}. Such magnetic parity violation is distinguished from the crystalline parity violation by the preserving symmetry; i.e., the former preserves the parity-time-reversal symmetry (\PT symmetry), the symmetry of combined \Pa and \T operations. The two types of parity violation affect the equilibrium and transport properties as well as the photocurrent response in a qualitatively different manner~\cite{Watanabe2018grouptheoretical,Hayami2018Classification}. A representative mechanism of the magnetic parity violation is the antiferromagnetic ordering of spin moments in a system hosting the specific crystal structure called a locally noncentrosymmetric system~\cite{Maruyama2012,Yanase2014zigzag,Hayami2014}. Supposing the antiferromagnetic order denoted by the zero N\'eel vector $\bm{Q}= \bm{0}$ in the locally noncentrosymmetric system, the ordered phase has neither the \Pa nor \T symmetries while it respects the \PT symmetry. In fact, the previous studies reported the photocurrent responses originating from the antiferromagnetic order~\cite{Zhang2019switchable,Holder2020consequences,fei2020giant,Ahn2020,watanabe2020chiral}.

There is an alternative way to cause the magnetic parity violation without requiring the spin degree of freedom. In this paper, we consider the spontaneous order of orbital anapole moments. When the microscopic looped-currents of electrons form the orbital magnetic moment $\bm{m}_\text{o}$, correspondingly, they may induce the anapole moment defined as $\sim \nabla \times \bm{m}_\text{o}$~\cite{Varma1997,Scagnoli2011-tg}. The anapole moment, which is also called the toroidal moment, is a vectorial quantity whose symmetry is the same as that evoked by the magnetic parity violation and may cause various emergent responses such as nonlinear optical responses. Recent studies have shown that the orbital anapole order is a potential candidate for the order parameter of the pseudo-gap phase in cuprates~\cite{Varma1997,Fauque2006,Li2008-hj,Pershoguba2013-jy,Pershoguba2014-dh,Zhao2017-ka} and the hidden-ordered phase in iridates~\cite{Zhao2016SHG,Jeong2017,Murayama2020}. Note that the spin anapole moment can arise from the spin magnetic moments as discussed in magnetoelectric materials~\cite{Van_Aken2007-zf,Spaldin2008-zj,Zimmermann2014-pv}. In the following section, the anapole order denotes the orbital anapole order.

In this paper, we investigate the photocurrent response induced by the ferroic anapole order. The photocurrent response is sensitive to the parity violation and may be applicable to identifying the anapole order in matters as the second harmonic generation is~\cite{Van_Aken2007-zf,Zimmermann2014-pv,Zhao2017-ka}. In fact, the photocurrent measurements have been performed to examine various quantum materials such as cuprates~\cite{Lim2020BSCCOphotocurrent} and $1T$-TiSe$_2$, a candidate material for a long-sought excitonic insulator~\cite{Xu2020-ra}. To demonstrate the photocurrent performance of the anapole-ordered system, we look into \sro{}. In \sro, a hidden order which may not be explained by the antiferromagnetic order has been observed, and the second harmonic generation~\cite{Zhao2016SHG} and polarized neutron diffraction experiments~\cite{Jeong2017} imply the anapole order. Furthermore, a bulk measurement~\cite{Murayama2020} recently shows that the direction of the ordered anapole moments is different from the previously proposed one~\cite{Zhao2016SHG,Jeong2017}. The photocurrent experiment may play a key role in unambiguously detecting the anapole order and elucidating the structure of anapole moments in \sro. 

The organization of this paper is as follows. In Sec.~\ref{Sec_symmetry_analysis}, we present the basic symmetry analysis of the current-ordered states and classify possible anapole-ordered states in \sro{}. The photocurrent responses allowed in anapole-ordered states are also classified. In Sec.~\ref{Sec_sro_photocurrent}, we introduce a realistic model for \sro{} consisting of the $t_{2g}$ orbitals of iridium ions and calculate the photocurrent response in the anapole-ordered state proposed in Ref.~\onlinecite{Murayama2020}. The obtained result indicates that the magnetic parity violation originating from the anapole order gives rise to the photocurrent response, and interestingly the response is comparable to that of a prototypical semiconductor. In Sec.~\ref{Sec_discussion_summary}, we discuss the dependence of the photocurrent on the light's polarization state for an experimental test and summarize this paper.

\section{Symmetry analysis of the loop-current ordered states}
\label{Sec_symmetry_analysis}

\begin{figure}[htbp]
        \centering
            \includegraphics[width=0.85\linewidth,clip]{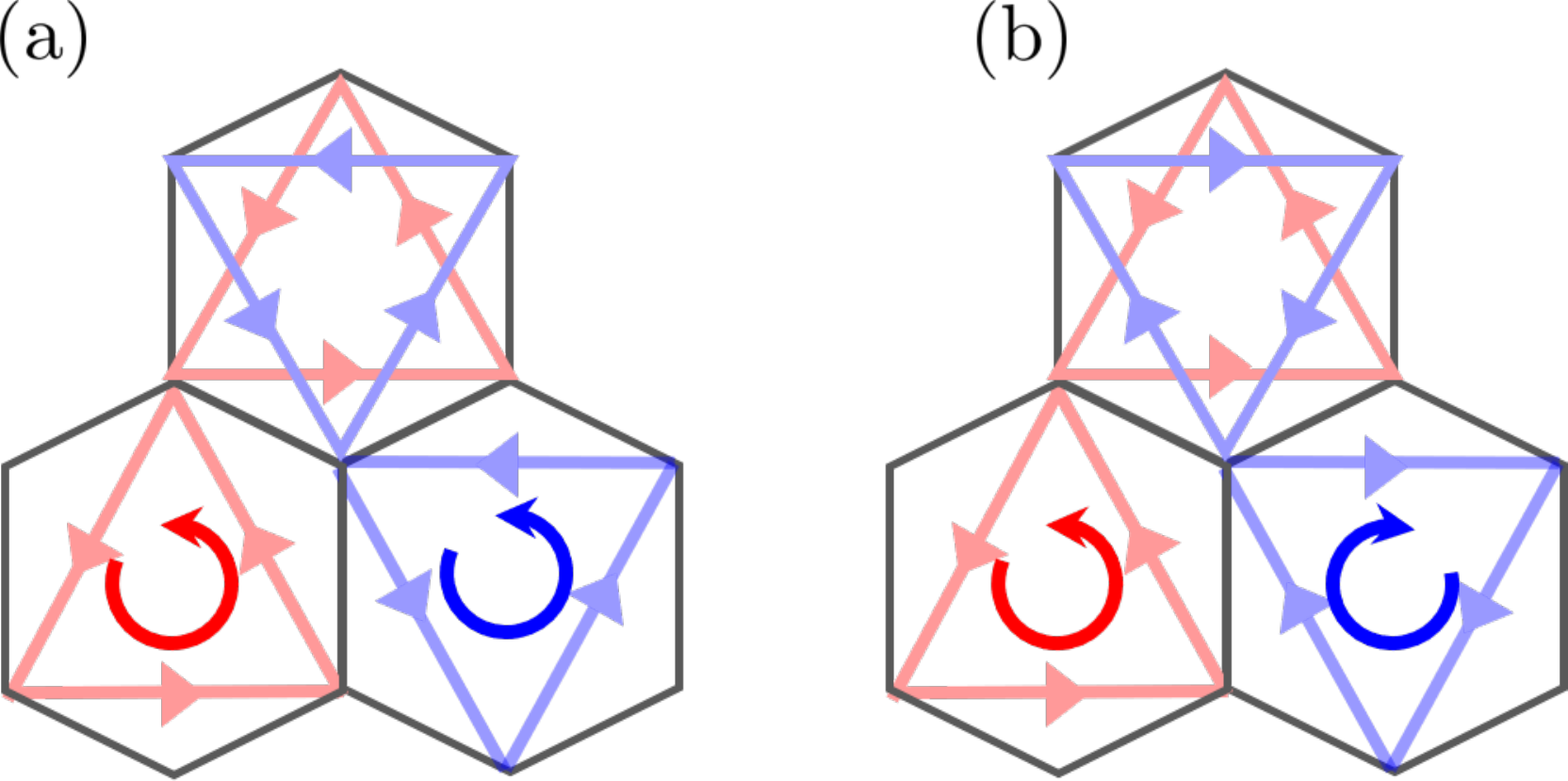}
        \caption{\T{}-violating current order in the honeycomb lattice. (a) \Pa{} symmetric (\PT{}-violating) current order originally introduced by Haldane \cite{Haldane1988-dm}. (b) \PT{} symmetric current order which breaks the \Pa{} symmetry.}
        \label{Fig_honeycomb_orbital_order}
        \end{figure}

A prototypical example of a current-ordered state is found in the quantum anomalous Hall system without net magnetization, namely, the Haldane model~\cite{Haldane1988-dm}. The Haldane model describes spinless fermions on the honeycomb lattice and comprises intra-sublattice complex hopping energy. Since the complex hopping is introduced to make a microscopic current flow and magnetic flux, the anomalous Hall effect can appear without the external magnetic field or spin degree of freedom. The current-ordered state preserving the \PT{} symmetry is similarly obtained, while the current ordering in the Haldane model breaks both of the \T{} and \PT{} symmetries (see Fig.~\ref{Fig_honeycomb_orbital_order})~\footnote{We note that the \PT symmetric current order in Fig.~\ref{Fig_honeycomb_orbital_order}(b) does not belong to the polar magnetic point group and cannot be denoted by the ferroic anapole order. It is labeled by the ordering of higher-rank magnetic multipole moments and toroidal multipole moments~\cite{Watanabe2018grouptheoretical,Hayami2018Classification}. }. 

The configuration of the current order preserving a specific symmetry can be systematically obtained by the representation theory as in magnetic structure determination~\cite{izyumov2012magnetic}.
Thus, the formulation for the current order can be generally performed in spinless systems; for instance, the \PT{} symmetric current order was proposed as a possible candidate for the order parameter of the pseudo-gap phase in cuprates and called loop-current order~\cite{Varma1997}. Its microscopic aspects can be captured by the so-called (spinless) Emery model which consists of the  Cu:~$d_{x^2-y^2}$ orbital and O:~$p_x,p_y$ orbitals~\cite{Emery1987}. 
 It is noteworthy that the loop-current order leads to the magnetic parity violation without the spin degree of freedom as in the variant of the Haldane model [Fig.~\ref{Fig_honeycomb_orbital_order}(b)]. 
As we show in the details of Sec.~\ref{Sec_sro_photocurrent}, the magnetic parity violation gives rise to the photocurrent response in the spinless systems with current order similarly to the spinful systems with spin order~\cite{Zhang2019switchable,watanabe2020chiral,Ahn2020}.

Here, we perform the symmetry analysis of \sro{}, a target material of this paper. The crystal structure is denoted by the space group $I4_1/a$ (No.~88)~\cite{Ye2015-vf}, while early studies reported the higher symmetry, $I4_1/acd$ (No.~142)~\cite{Crawford1994}. The difference, however, is only the slight change in the volume of the IrO$_6$ octahedra. Thus, we below consider the latter high-symmetric crystal structure for simplicity. Note that the result of symmetry analysis is similar to the case of the former low-symmetric crystal structure.

\begin{figure}[htbp]
    \centering
        \includegraphics[width=0.90\linewidth,clip]{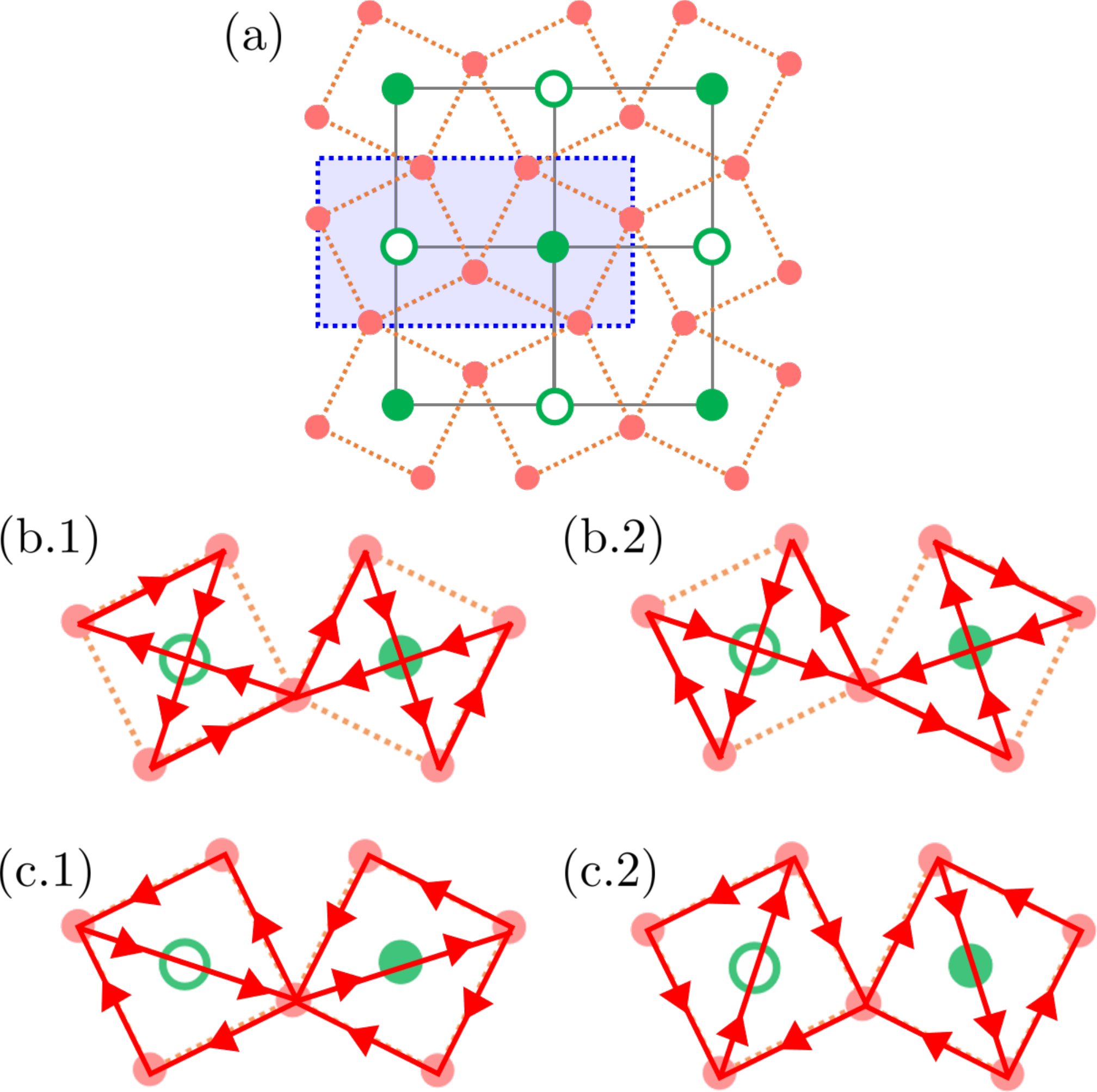}
    \caption{(a) Crystal structure of \sro{} in the two-dimensional $xy$ plane. The green-colored open and filled circles represent two sublattices of iridium sites. The pink-filled circles denote the oxygen sites.
    The blue-shaded region represents a unit cell of Ir-O square network.
    (b) The \PT symmetric current order preserving the $[110]$-axis two-fold rotation ($E_u^- \Braket{d}$ order). The red directed lines denote the microscopic electric current. (c) The \PT symmetric current order preserving the $[100]$-axis two-fold rotation [$E_u^- \Braket{x}$ order].
    Two candidates for the $E_u^- \Braket{d}$ order are shown in (b.1) and (b.2). The same applies to (c.1) and (c.2) for the $E_u^- \Braket{x}$ order.}
    \label{Fig_iro_current_order_configs}
    \end{figure}

Although \sro{} comprises four sublattices of Ir sites, two of them almost independently form the Ir-O two-dimensional net [Fig.~\ref{Fig_iro_current_order_configs}(a)] with the quasi-two-dimensional electron bands.
Let us consider the \PT{} symmetric current order in the two-dimensional layer. Supposing the current order at the bonds between the nearest neighbor Ir$-$O and O$-$O sites, we obtain the two possible patterns as shown in Fig.~\ref{Fig_iro_current_order_configs}(b,\,c).
We here assume that the current order respects the Kirchhoff law at each site.
Figures~\ref{Fig_iro_current_order_configs}(b.1) and~\ref{Fig_iro_current_order_configs}(b.2) show two-fold patterns of the current order having the polar axis along the $[110]$-axis.
On the other hand, the patterns shown in Figs.~\ref{Fig_iro_current_order_configs}(c.1) and (c.2) have the polar axis along the $[100]$-direction.
Here we consider the hidden order in \sro{} which does not break the translational symmetry. Thus, the current pattern in a unit cell is depicted.
The current patterns correspond to the ferroic ordering of the anapole moments aligned along the $[110]$-axis and the $[100]$-axis.
In the group-theoretical terminology, the current patterns are denoted by the $E_u$ representation of a tetragonal point group $D_{4h}$. Since the representation $E_u$ is two-dimensional, we can find pairs of two independent patterns. For instance, Fig.~\ref{Fig_iro_current_order_configs}(c.1) is paired with that rotated by $\pi/2$ in the $xy$-plane. The paired configurations can be labeled by $E_u \Braket{x}$ and $E_u \Braket{y}$ which indicate that the two-fold rotation symmetries along the $x~([100])$-axis and $y~([010])$-axis are preserved, respectively. In a similar manner, we label the configurations in Fig.~\ref{Fig_iro_current_order_configs}(b) by $E_u \Braket{d}$ due to the preserved rotation symmetry along the diagonal axis, that is, $[110]$-axis.
We also label the irreducible representation by the parity under the time-reversal operation as $\Gamma^\pm$, e.g., the configurations in Fig.~\ref{Fig_iro_current_order_configs}(b) are \T{}-odd and denoted such as $E_u^- \Braket{d}$.

The bond order in Fig.~\ref{Fig_iro_current_order_configs}(b.1) agrees with the configuration proposed in Ref.~\onlinecite{Jeong2017}. According to recent bulk measurements~\cite{Murayama2020}, on the other hand, the possible anapole order is labeled by the crystal symmetry in which the $[100]$-axis rotation symmetry is preserved. The corresponding configurations are shown in Fig.~\ref{Fig_iro_current_order_configs}(c). Note that the anapole orders depicted in Figs.~\ref{Fig_iro_current_order_configs}(b.2) and~\ref{Fig_iro_current_order_configs}(c.2) can be \PT{} symmetric parity-violating order only when the IrO$_6$ octahedra rotations are considered. These current patterns give rise to only the microscopic translation symmetry breaking in the absence of the octahedra rotations. 
We also note that the three-dimensional pattern of the current order can be obtained by the simple stacking of the configuration in the two-dimensional net.
The inversion center of the two-dimensional net in Fig.~\ref{Fig_iro_current_order_configs}(a) can be taken at the Ir sites and differs from that of the real crystal structure of \sro{}. This difference, however, does not affect the present symmetry argument of the anapole-order-induced magnetic parity violation, while the local \Pa{} symmetry breaking is essential for that induced by the spin order~\cite{Matteo2016}.

\begin{figure}[htbp]
    \centering
        \includegraphics[width=0.85\linewidth,clip]{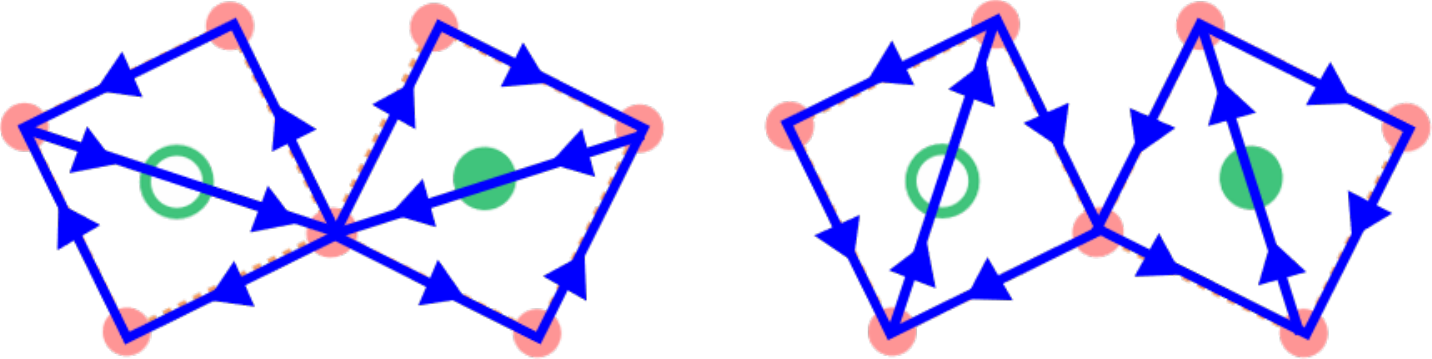}
    \caption{Two-fold candidate patterns of the parity-violating spin current order in \sro{} [$E_u^+ \Braket{x}$ order].
    The two-fold rotation symmetry along the $[100]$ axis is preserved. The blue directed lines denote the spin current polarized in the $+\hat{z}$ direction.}
    \label{Fig_sro_spincurrent_eux}
    \end{figure}

We comment on the spin current order. 
When the spin-polarized currents are placed at the bonds of Fig.~\ref{Fig_iro_current_order_configs}, the obtained \textit{spin current} ordered state preserves the \T{} symmetry instead of the \PT{} symmetry. The \T{}-breaking/preserving correspondence between the spin current and current ordered states can also be found in the Haldane model~\cite{Haldane1988-dm} and the Kane-Mele model for the quantum spin-Hall insulator~\cite{Kane2005}. 
With the $z$-polarized spin current, we can also obtain the parity-violating spin current order and show an example in  Fig.~\ref{Fig_sro_spincurrent_eux}. The obtained spin-polarized bond order is consistent with the experimental results in Ref.~\onlinecite{Murayama2020}, although it may not be consistent with the polarized neutron diffraction experiment~\cite{Jeong2017}.

Following the general classification based on \T{} and \PT{} symmetries~\cite{watanabe2020chiral}, we obtain the photocurrent responses and components of the photocurrent coefficients as described in Table.~\ref{Table_photocurrent_sro_bond_order_corespondence}.
Owing to the parity under the \T{} operation, the current order is denoted by the representation $\Gamma^-$, while the spin current order by $\Gamma^+$.
The notation for the photocurrent coefficients $\eta^{\mu;\nu\lambda}$ and $\kappa^{\mu\nu}$ will be described in Sec.~\ref{Sec_sro_photocurrent}. We note that the photocurrent response arising from the planar current order can be distinguished from the surface effect by the symmetry; i.e., the surface effect is characterized by the irreducible representation $A_{2u}^+$ of the point group $D_{4h}$, while the current and spin current order in Figs.~\ref{Fig_iro_current_order_configs} and~\ref{Fig_sro_spincurrent_eux} are labeled by the irreducible representation $E_u^\pm$. Thus, the allowed photocurrent coefficients have the different symmetry.

		\begin{table}[htbp]
        \caption{Table of the photocurrent tensor in the (spin) current-ordered states.
        Notation for $\eta^{\mu;\nu\lambda}$ and $\kappa^{\mu\nu}$ is described in Eq.~\eqref{eq_photocurrent_formula}. $E_u$ is the irreducible representation of the point group $D_{4h}$. The superscript $\pm$ denotes the parity under the \T{} operation. $\Braket{x}$ and $\Braket{d}$ respectively indicate the two-fold rotation symmetry along the $[100]$ and $[110]$-axes. The $E_u^- \Braket{d}$ and $E_u^- \Braket{x}$ correspond to the current order in Figs.~\ref{Fig_iro_current_order_configs}(b) and~\ref{Fig_iro_current_order_configs}(c), respectively. 
        The response tensors in the parenthesis are allowed only in the three-dimensional system.}
        \label{Table_photocurrent_sro_bond_order_corespondence}
        \centering
        \begin{tabular}{l|ll} \hline\hline
            $\Gamma$&$\eta^{\mu;\mu\lambda}$&$\kappa^{\mu\nu}$\\  \hline
            $E_u^\pm \Braket{x}$&
            $\eta^{x;xx},~\eta^{y;xy},~\eta^{x;yy},$&
            $\kappa^{yz},~\left( \kappa^{zy} \right)$\\
            &$~\left( \eta^{z;zx},~\eta^{x;zz} \right)$&\\
            $E_u^\pm \Braket{d}$&
            $\eta^{x;xx}=\eta^{y;yy},~\eta^{y;xy}=\eta^{x;yx},$&
            $\kappa^{yz}=\kappa^{xz},$\\
            &$\eta^{x;yy}=\eta^{y;xx},$&$\left( \kappa^{zy}=\kappa^{zx} \right)$\\
            &$\left( \eta^{z;zx}=\eta^{z;zy},~\eta^{x;zz}=\eta^{y;zz} \right)$&\\
            \hline\hline
        \end{tabular}
        \end{table}
In the above discussions, we classified the possible (spin) current order in the hidden-ordered phase without magnetic order. \sro{} undergoes antiferromagnetic order below the hidden-order transition temperature, and thus, we take into account the antiferromagnetic order in the following. The magnetic structure is characterized by the non-zero N\'eel vector $\bm{Q} = [111]$~\cite{Kim2009,Dhital2013,Ye2015-vf}. Owing to the magnetic doubling of the unit cell, the magnetic point group symmetry is reduced to
		\begin{equation}
        mmm1',
        \end{equation} 
when we do not consider the current order and weak ferromagnetic component.
The two-fold rotation symmetries are in the $[110]$, $[\bar{1}10]$, and $[001]$-axes.

In the microscopic calculation below, we take into account the coexisting current order and antiferromagnetic order and study the low-temperature phase. 
Although the symmetry classification in Table~\ref{Table_photocurrent_sro_bond_order_corespondence} is modified by the admixture with the antiferromagnetic order, we investigate the photocurrent coefficients allowed in the current-ordered state irrespective of the presence of the antiferromagnetic order. 
Although this paper does not elaborate on a quantitative relation between the ordered microscopic anapole moments and photocurrent response, recent theoretical developments may resolve the issue~\cite{Shitade2018,Gao2018orbitalMQM}.

\section{Photocurrent response in Iridate}
\label{Sec_sro_photocurrent}
Based on the symmetry analysis, we construct the model of \sro and investigate the photocurrent response arising from the \PT{} symmetric current order. With the minimal model introduced in Sec.~\ref{Sec_sro_model}, we calculate the linearly and circularly-polarized-light-induced photocurrent responses and propose the experimental detection of the current order in Sec.~\ref{Sec_sro_LC_photocurrents}.

\subsection{Model}
\label{Sec_sro_model}
We adopt Ir:~$t_{2g}$ orbitals as the model space which captures the electronic structure near the Fermi level and construct the two-dimensional three-orbital model. According to the first-principles calculation~\cite{HiroshiWatanabe2010}, the Hamiltonian reads
		\begin{equation}
        H = \sum_{\bk} \bm{c}_{\bk}^\dagger \text{H}_{\bk} \bm{c}_{\bk}, \label{total_Hamiltonian}
        \end{equation}
where basis for creation operators $c_{\bk;a,\tau,\sigma}^\dagger$ are spanned by the orbital ($a=yz,zx,xy$), sublattice ($\tau=A,B$), and spin ($\sigma=\pm$) degrees of freedom. When the creation operators are distinguished by the sublattice degree of freedom as $\bm{c}_{\bk}^\dagger = \left( \bm{c}_{\bk;A}^\dagger,\bm{c}_{\bk;B}^\dagger  \right)$, each creation operator vector labeled by the sublattice $\tau=A,B$ is written as 
		\begin{align}
           \bm{c}_{\bk;\tau}^\dagger
                =  &\Bigl( c_{\bk;yz,\tau,\uparrow}^\dagger,c_{\bk;yz,\tau,\downarrow}^\dagger,c_{\bk;zx,\tau,\uparrow}^\dagger,\notag \\
                    &~~c_{\bk;zx,\tau,\downarrow}^\dagger,c_{\bk;xy,\tau,\uparrow}^\dagger,c_{\bk;xy,\tau,\downarrow}^\dagger  \Bigr).
        \end{align}
The momentum-resolved Hamiltonian $\text{H}_{\bk}$ is given by
		\begin{equation}
        \text{H}_{\bk} 
            = \text{diag} (\epsilon_{yz},\epsilon_{zx},\epsilon_{xy})\, \sigma_0\tau_0 + \text{H}_\text{SOC} + \text{H}_\text{AFM} +\text{H}_\text{CO}.
        \end{equation}
The Pauli matrices $\bm{\sigma}$ and $\bm{\tau}$ denote the spin and sublattice degrees of freedom. The kinetic energies for each orbital are defined by
		\begin{align}
            &\epsilon_{yz} = -2t_5 \cos{k_x}-  2t_4\cos{k_y},\\
            &\epsilon_{zx} = -2t_4 \cos{k_x}-  2t_5\cos{k_y},\\
            &\epsilon_{xy} = -2t_1 \left(  \cos{k_x}+\cos{k_y} \right)-4t_2 \cos{2k_x}\cos{2k_y} +\delta.
        \end{align}
Since the IrO$_6$ octahedrons are tilted in the basal plane, the crystal field correction $\delta$ is included in the kinetic energy of the $d_{xy}$ orbital. The spin-orbit coupling is introduced by the LS coupling,
		\begin{equation}
            \text{H}_\text{SOC} 
            =
            \frac{\lambda_\text{SOC}}{2}
            \begin{pmatrix}
                &   &   i&   &   & -1  \\
                &   &   &   -i&   1&   \\
                -i&   &   &   &   &i   \\
                &   i&   &   &   i&   \\
                &   1&   &   -i&   &   \\
                -1&   &   -i&   &   &   
            \end{pmatrix} \tau_0.
        \end{equation}
The parameters are chosen to be consistent with a LDA+SOC calculation as follows
		\begin{align}
            &(t_1,t_2,t_3,t_4,t_5) = (0.36,0.18,0.09,0.37,0.06),\\
            &\delta = -0.36,~\lambda_\text{SOC} = 0.37.
        \end{align}
The unit is an electron volt (eV). Since the strong spin-orbit coupling of Ir atoms almost surpasses the kinetic energy, the $t_{2g}$ manifold is split into the $j=1/2$ singlet and $j=3/2$ quartet. It is therefore convenient to introduce the basis denoted by the effective angular momentum $j = 1/2,3/2$~\cite{Kim2008SOCMott}. In particular, the electronic structure near the Fermi energy mainly consists of the $j=1/2$ orbital,
		\begin{equation}
        \ket{j=\frac{1}{2},\sigma} = \frac{1}{\sqrt{3}} \left(  \ket{d_{xy},\sigma} +\sigma \ket{d_{yz},\sigma} +i \ket{d_{zx},-\sigma} \right).
        \end{equation}
        
The remaining terms $\text{H}_\text{AFM}$ and $\text{H}_\text{CO}$ represent the molecular fields of the antiferromagnetic order and the current order shown in Fig.~\ref{Fig_iro_current_order_configs}(c.1). The model is a one-body Hamiltonian, although \sro{} is known to be the Mott insulator due to the half-filled $j=1/2$ band~\cite{Kim2008SOCMott}. Thus, the model may be applicable only to the low-temperature antiferromagnetic phase. The insulating antiferromagnetic state is described by the mean-field model when the quantum fluctuation is not significant.
The antiferromagnetic molecular field $\text{H}_\text{AFM}$ is introduced as follows. The magnetic moments arise from the $j=1/2$ orbitals, and hence the molecular field is written by
		\begin{equation}
        \sum_{\bk ss'} d_{\bk; \tau s}^\dagger \bm{h} \cdot \bm{\sigma}_{ss'} {(\tau_z)}_{\tau \tau'} d_{\bk;\tau' s'},\label{AFM_molecular_field}
        \end{equation}
in which we introduce the creation and annihilation operators $d_{\bk;s \tau}$ for the $j=1/2$ orbital labeled by the sublattice $\tau$ and spin $s$. The molecular field is taken as $\bm{h} \parallel [110]$ to be consistent with the recent magnetic structure determination~\cite{Kim2009,Dhital2013,Ye2015-vf}. A small ferromagnetic component due to the canted antiferromagnetic order is neglected for simplicity~\cite{HiroshiWatanabe2010}. The directional current order is built upon the $j=1/2$ orbitals as
    \begin{equation}
    \sum_{\bk s} -2u \sin{k_x} d_{\bk;B s}^\dagger  d_{\bk;A s} + h.c..\label{current_order_molecular_field}
    \end{equation}
The molecular fields $\text{H}_\text{AFM}$ and $\text{H}_\text{CO}$ in the spin-orbital basis are obtained by applying the unitary transformation to Eqs.~\eqref{AFM_molecular_field} and~\eqref{current_order_molecular_field}.
The molecular field of the antiferromagnetic order is taken as $h=|\bm{h}|=0.368$ to reproduce the electronic gap $\sim \mr{0.07}{eV}$~\cite{Kim2008SOCMott}.
The current order is essential for the magnetic parity violation, and the photocurrent response shows the monotonous increase with increasing $u$. In this paper, the adopted value $u=0.03$ is comparable to the transition temperature of the hidden order~\cite{Murayama2020}.
For quantitative estimation of the response coefficients, we also use the lattice constants reported in Ref.~\onlinecite{Ye2015-vf}.

Corresponding to the current order $E_u^- \Braket{x}$ in Table~\ref{Table_photocurrent_sro_bond_order_corespondence}, the magnetic crystal symmetry is transformed as 
		\begin{equation}
        4/mmm1' \rightarrow m'mm.\label{current_order_symmetry_reduction}
        \end{equation}
The ordered phase respects the two-fold rotation symmetry along the $[100]$-axis. When the antiferromagnetic order is taken into account, the symmetry is further reduced as 
    \begin{equation}
        m'mm \rightarrow 2'/m.\label{AFM_current_order_symmetry_reduction}
    \end{equation}
The mirror plane is $(001)$. Importantly, the magnetic symmetries in both cases respect neither \Pa{} nor \T{} symmetry while they have the \PT{} symmetry. 

We validate the adopted model by the calculation of the optical attenuation coefficient. Under the irradiating lights with energy $\hbar \Omega$, the optical attenuation coefficient is given by
    \begin{equation}
    \varepsilon^{\mu\nu}_\text{att} 
        = \pi q^2 \int \frac{d\bm{k}}{\left( 2\pi \right)^2} \sum_{a, b} \mathcal{A}^\mu_{ab} \mathcal{A}^\nu_{ba} f_{ab} \delta (\hbar \Omega -\epsilon_{ba}),\label{attenuation_coefficient}
    \end{equation}
where the electron charge is $q = -e <0$ and $a,b$ are the band indices~\cite{Sipe2000secondorder,Souza2008}. The Bloch equation is given by
    \begin{equation}
        H (\bm{k}) \ket{u_{a} (\bm{k})} = \epsilon_{\bm{k}a} \ket{u_{a} (\bm{k})},
    \end{equation}
where $H (\bm{k})$ is the Bloch Hamiltonian obtained from Eq.~\eqref{total_Hamiltonian} and $\ket{u_{\bk a}}$ is the (periodic) Bloch state. Accordingly, we define $\epsilon_{ab} = \epsilon_{\bk a}-\epsilon_{\bk b}$ and $f_{ab} = f \left(  \epsilon_{\bk a} \right) - f \left(  \epsilon_{\bk b} \right)$ where $f \left(  x \right) = 1 / (1+ \exp{\left[ \left(   x-\mu \right)/T \right]})$ is the Fermi distribution function parametrized by the chemical potential $\mu$ and temperature $T$. 
We also define the interband Berry connection $\bce^\mu = i \Braket{ u_{\bk a}| \partial_\mu u_{\bk b} }$ which is non-zero when the Bloch states $\ket{u_{\bk a}},\ket{u_{\bk b}}$ have different energy eigenvalues~\cite{watanabe2020chiral}.  Throughout this paper, we take the chemical potential $\mu = \mr{0.522}{eV}$ so as to be consistent with the insulating behavior of \sro{}. The optical conductivity is obtained by the attenuation coefficient as
		\begin{equation}
        \text{Re}\,\sigma^{\mu\nu} (\Omega) = \Omega \, \varepsilon^{\mu\nu}_\text{att} (\Omega).
        \end{equation}
For a comparison we also calculate the frequency dependence of the joint density of states written by
\begin{equation}
    J_d \left( \Omega \right)= \frac{1}{V} \sum_{a\neq b}\int d\bm{k} \delta (\hbar \Omega - \epsilon_{ba})f_{ab},\label{joint_density_of_states}
    \end{equation}
where $V$ is the volume of the first Brillouin zone. The joint density of states counts the electron-hole pairs created by the irradiating lights. For numerical calculations, we introduce a scattering rate $\gamma$.
The scattering rate phenomenologically introduces the broadening of the spectrum, which may arise from the electron correlation and disorder scattering.
Correspondingly, the delta function comprised in Eqs.~\eqref{attenuation_coefficient} and~\eqref{joint_density_of_states} is approximated by the Lorentian function
    \begin{equation}
        \mathcal{L} \left( x \right) = \frac{\gamma}{\pi}\frac{1}{x^2 + \gamma^2}.
    \end{equation}
The numerical calculations are performed in a $N$-discretized Brillouin zone.

\begin{figure}[htbp]
            
        \centering
        \includegraphics[width=0.98\linewidth,clip]{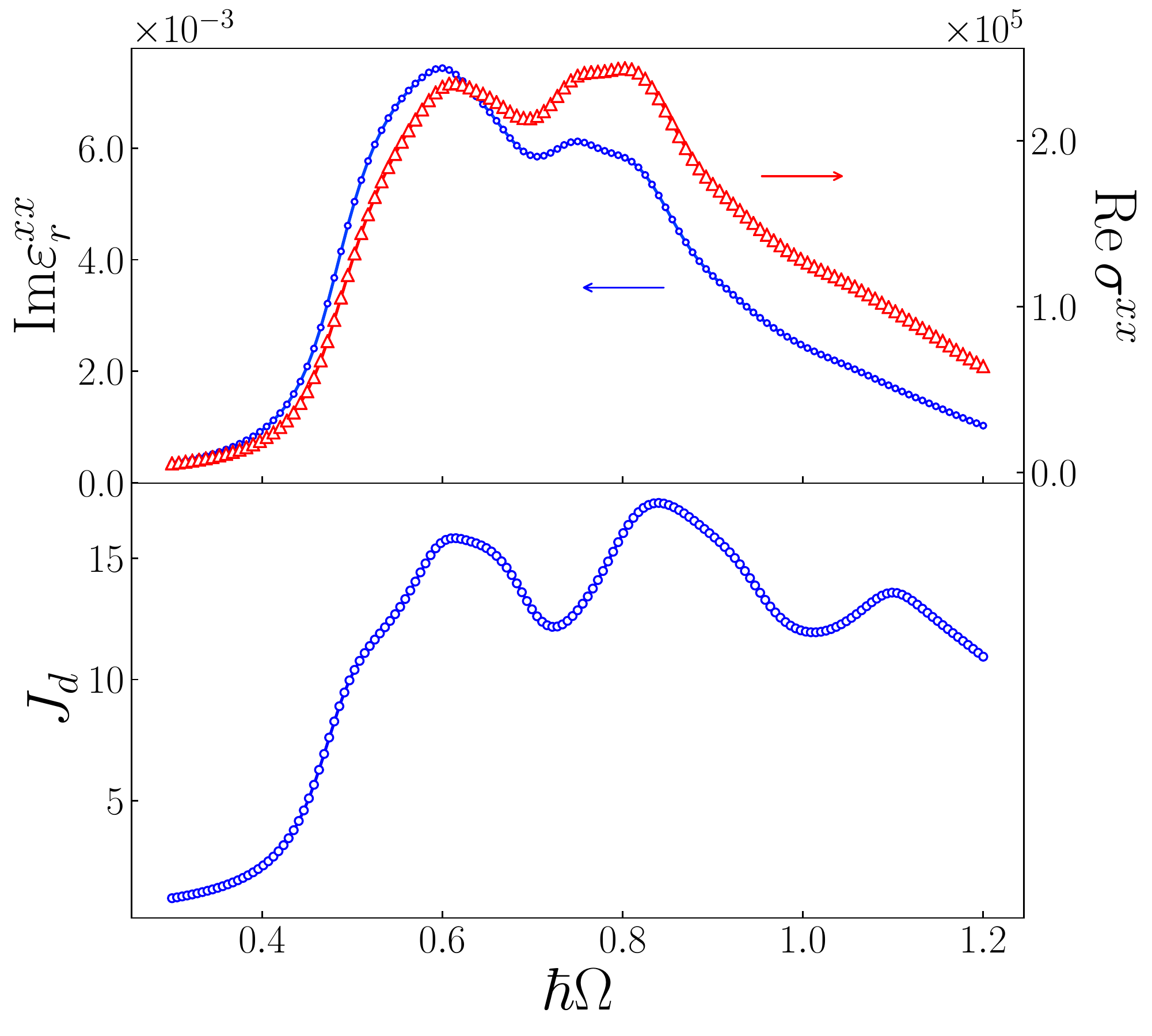}
        \caption{Frequency dependence of the (dimensionless) optical attenuation coefficient $\text{Im}\, \varepsilon^{xx}_r$ (blue-colored), optical conductivity $\text{Re}\, \sigma^{xx}$ $\Omega^{-1}\cdot$m$^{-1}$ (red-colored), and joint density of states $J_d$ eV$^{-1}$. The unit of the transverse axis $\hbar \Omega$ is electron volt (eV). The optical attenuation coefficient in Eq.~\eqref{attenuation_coefficient} is converted into the dimensionless one $\varepsilon^{xx}_r $ by using the vacuum permittivity  $\varepsilon_0 = \mr{8.85 \times 10^{-12} }{F\cdot m^{-1}}$ and the thickness of the two-dimensional net $ l = \mr{6.5}{\AA}$. The optical conductivity in the three-dimensional compound is also calculated by using the thickness $l$. The parameters $N = 1200^2$, $T = \mr{0.001}{eV}$, and $\gamma = \mr{0.04}{eV}$ are adopted.}
        \label{Fig_attenuation_jdos_plot}
        \end{figure}

The results of the calculation are shown in Fig.~\ref{Fig_attenuation_jdos_plot}.
Here, we adopt a moderate scattering rate $\gamma= \mr{0.04}{eV}$ to reproduce the peak structure observed in the optical conductivity measurement~\cite{Kim2008SOCMott}.
In particular, the intense excitation at $\sim\mr{0.5}{eV}$ is attributed to the transition between the $j=1/2$ orbitals and plays an important role in the optical phenomena in the spin-orbit-coupled system, \sro{}~\cite{HiroshiWatanabe2010}.
The obtained model Hamiltonian therefore captures the optical property of \sro{}, although it may be insufficient in the higher frequency regime, $ \hbar \Omega \gtrsim \mr{2.0}{eV}$.

\subsection{Linear and circular photocurrent responses}
\label{Sec_sro_LC_photocurrents}

The formula for the photocurrent response is given by
		\begin{align}
        J_\text{PC}^\mu 
            &= \int \frac{d \Omega}{2\pi}  \sigma^{\mu;\nu\lambda} (0;-\Omega,\Omega) E^{\nu} (-\Omega) E^{\lambda} (\Omega),\\
            &=  \int \frac{d \Omega}{2\pi} \left[  \eta^{\mu;\nu\lambda} (\Omega)  L^{\nu\lambda}(\Omega)  + \kappa^{\mu\nu} (\Omega)  F^{\nu}(\Omega) \right],
        \label{eq_photocurrent_formula}
        \end{align}
with the second-order nonlinear conductivity $\sigma^{\mu;\nu\lambda}$ and the photo-electric field $\bm{E} \left(  \Omega \right)$. The symmetrized and antisymmetized conductivities are respectively introduced by $\eta^{\mu;\nu\lambda} =\sigma^{\mu;\nu\lambda}/2 + \sigma^{\mu;\lambda\nu} /2$ and $\kappa^{\mu\tau} = i \epsilon_{\nu\lambda\tau } \sigma^{\mu;\nu\lambda}$. We also define
        \begin{align}
            &L^{\nu\lambda} (\Omega)= \text{Re} \left[  E^{\nu} (\Omega) (E^{\lambda} (\Omega) )^\ast \right],\\
            &\bm{F} (\Omega) = \frac{i}{2} \bm{E}(\Omega) \times \bm{E}^\ast (\Omega),	
        \end{align}
which are finite under the linearly and circularly-polarized lights, respectively~\cite{Sturman1992Book}. The symmetry of the photocurrent response tensor is determined by the crystal symmetry (unitary symmetry), whereas it is not subject to the strong constraint from the magnetic symmetry (anti-unitary symmetry), which includes the time-reversal operation. The magnetic symmetry, however, plays an essential role in determining the mechanism of photocurrent. 
For instance, let us consider the shift current mechanism, one of the well-known mechanisms for photocurrent generation~\cite{Kraut1981Photovoltaiceffect,Sipe2000secondorder}. The photocurrent derived from the shift current mechanism is induced only by the linearly-polarized light in the \T symmetric system, while it is caused only by the circularly-polarized light in the \PT symmetric system~\cite{Ahn2020,watanabe2020chiral}. These photocurrents induced by the linearly and circularly-polarized lights are called shift current and gyration current, respectively~\cite{watanabe2020chiral}.

In this study, we examine the \PT{} symmetric anapole order in the insulating \sro{}. Following the general symmetry argument~\cite{watanabe2020chiral}, the photocurrent mechanism is identified as the magnetic injection current and gyration current, which respectively appear under the linearly and circularly-polarized lights~\cite{Zhang2019switchable,Holder2020consequences,Ahn2020,watanabe2020chiral}. Other contributions are classified into the Fermi-surface effect and hence excluded from the present study focused on the insulating state. The formulas are written by
    \begin{align}
    \eta^{\mu;\nu \lambda}_\text{inj} 
        &= \lim_{\omega \rightarrow 0} \frac{- i\pi  q^3}{\hbar  \omega}\int \frac{d\bm{k}}{\left( 2\pi \right)^d} \notag \\
        &\times \sum_{a\neq b} \left( v^\mu_{aa} - v^\mu_{bb} \right) \left(  \bce^{\nu}_{ab} \bce^{\lambda}_{ba} + \bce^{\lambda}_{ab} \bce^{\nu}_{ba}  \right) f_{ab} \delta (\hbar \Omega  - \epsilon_{ba} ).\label{magnetic_injection_current_bare}
    \end{align}
for the magnetic injection current and 
    \begin{align}
    \kappa^{\mu\nu}_\text{gyro}  
        &= \frac{\pi q^3}{\hbar} \int \frac{d\bm{k}}{\left( 2\pi \right)^d} \notag \\
        &\times \sum_{a\neq b}\epsilon_{\nu\lambda\tau } \text{Re}\, \left( \left[ \ud_\mu  \bce^\lambda \right]_{ab} \bce^\tau_{ba}\right) f_{ab} \delta (\hbar \Omega - \epsilon_{ba}),\label{gyration_current_formula}
    \end{align}
for the gyration current. We introduce the velocity operator $v^\mu$ and the covariant derivative $\ud$ which acts on the physical quantity $O$ as
    \begin{equation}
    [\ud_\mu O]_{ab} = \partial_\mu O_{ab} - i \left( \sum_c \alpha^\mu_{ac} O_{ca} - \sum_c O_{ac}\alpha^\mu_{cb}\right).\label{UN_gauge_covariant_derivative}
    \end{equation}
in the Bloch representation~\cite{watanabe2020chiral}. We also introduce the U($n$) Berry connection $\alpha^\mu$, where $n$ denotes the degeneracy at each crystal momentum $\bk$. We usually take $n=1$ for spinless systems and $n=2$ for spinful systems due to the \PT-ensured Kramers degeneracy.
Note that we can take $n=1$ in the shift current formula for \T{} symmetric noncentrosymmetric systems even with the spin degree of freedom, since the parity breaking and spin-orbit interaction lift the Kramers degeneracy at each $\bk$~\cite{Sipe2000secondorder}.
Since we work on the spinful model of \sro{}, the U(2)-covariant derivative is adopted. Numerical calculations are performed with a phenomenological scattering rate $\gamma$. Correspondingly, the magnetic injection current is regularized as
\begin{align}
    \eta^{\mu;\nu \lambda}_\text{inj} 
        &= \frac{- \pi  q^3}{\gamma}\int \frac{d\bm{k}}{\left( 2\pi \right)^d} \notag \\
        &\times \sum_{a\neq b} \left( v^\mu_{aa} - v^\mu_{bb} \right) \left(  \bce^{\nu}_{ab} \bce^{\lambda}_{ba} + \bce^{\lambda}_{ab} \bce^{\nu}_{ba}  \right) f_{ab}  \mathcal{L} \left( \hbar \Omega  - \epsilon_{ba} \right).\label{magnetic_injection_current_bare_with_scattering}
    \end{align}
Similarly, the gyration current formula is modified by substituting the Lorentian function for the delta function in Eq.~\eqref{gyration_current_formula}.

With the symmetry reduction due to the anapole order [Eq.~\eqref{current_order_symmetry_reduction}], the magnetic injection current coefficients $\eta_\text{inj}^{x;xx},\eta_\text{inj}^{x;yy},\eta_\text{inj}^{y;xy}$ and gyration current coefficient $\kappa_\text{gyro}^{yz}$ are allowed in the two-dimensional system (see Table~\ref{Table_photocurrent_sro_bond_order_corespondence}). We confirmed that the numerical results of the photocurrent responses are consistent with the symmetry analysis. The antiferromagnetic order further reduces the magnetic symmetry as in Eq.~\eqref{AFM_current_order_symmetry_reduction} and hence may give rise to other photocurrent responses such as that denoted by $\eta^{y;yy}$. Such components, however, do not appear in our model. Since the antiferromagnetic order does not break the \Pa{} symmetry and plays a minor role in the photocurrent response, it is expected that the photocurrent responses are not significantly influenced by the antiferromagnetic order. Supporting this statement, the numerical results of the photocurrent response are almost the same as those when the antiferromagnetic molecular field is set to be $\bm{h} \parallel [100]$. An essential role of the antiferromagnetic order is making the system insulating, whereas the photocurrent responses are mainly due to the anapole order.

\begin{figure}[htbp]
    \centering
    \includegraphics[width=0.95\linewidth,clip]{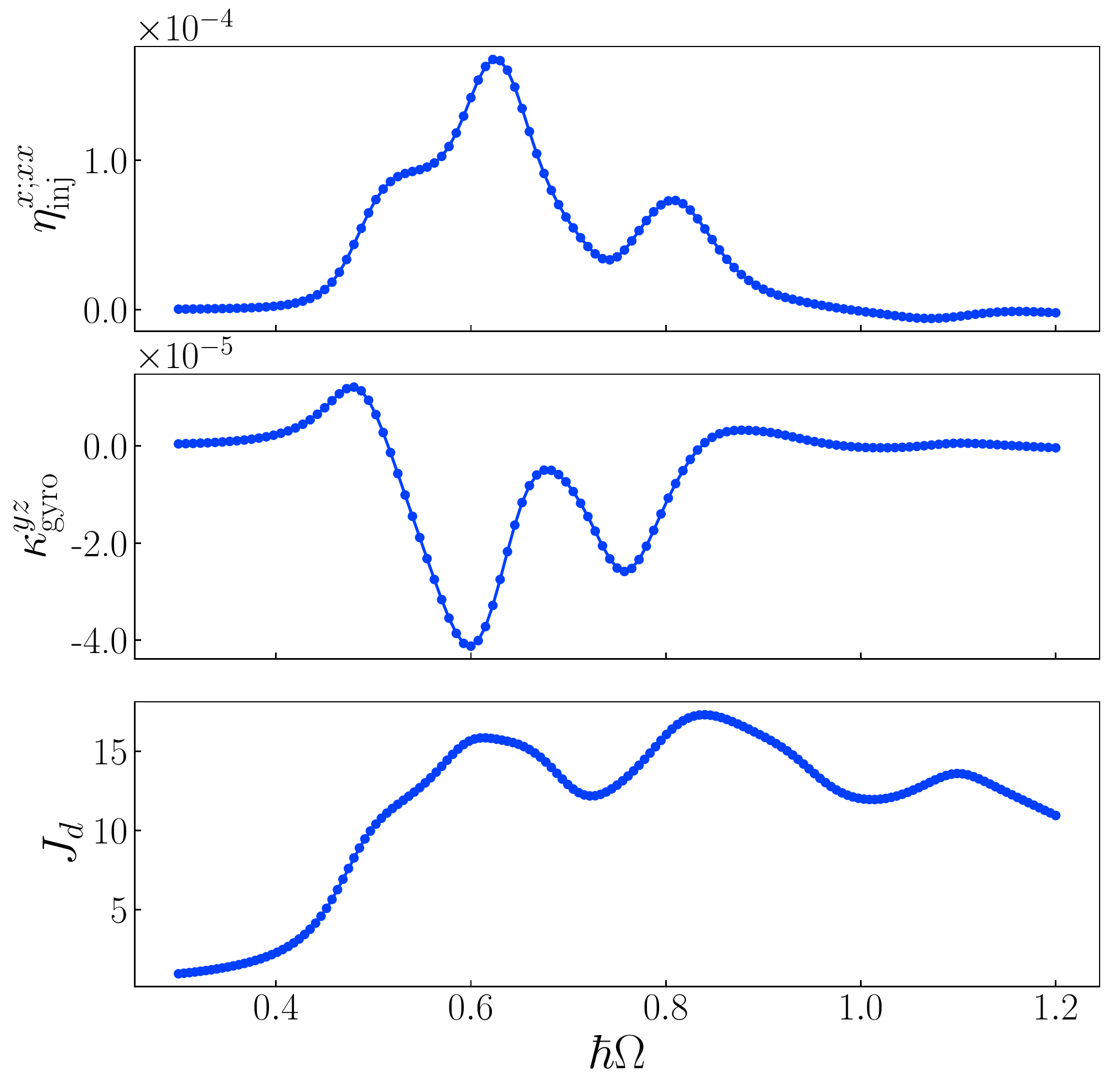}
    \caption{Frequency dependence of the magnetic injection current $\eta_\text{inj}^{x;xx}$ [A$\cdot$V$^{-2}$] and the gyration current $\kappa_\text{gyro}^{yz}$ [A$\cdot$V$^{-2}$]. The joint density of states $J_d$ [eV$^{-1}$] is replotted for comparison. We show the other components of the magnetic injection current in Appendix~\ref{Appendix_MIC}. The unit of the transverse axis is electron volt (eV). The parameters are the same as in Fig.~\ref{Fig_attenuation_jdos_plot}.}
    \label{Fig_gyration_injection_comparison}
    \end{figure}

The numerical calculations of the injection current and gyration current responses are shown with the plot of the joint density of states in Fig.~\ref{Fig_gyration_injection_comparison}. It is clearly shown that the anapole order gives rise to the photocurrent responses. As in the case of the optical attenuation coefficient in Eq.~\eqref{attenuation_coefficient}, the peak structure in the frequency dependence of the photocurrent responses roughly coincides with that of the joint density of states except for the higher frequency regime where the optical transition amplitude is negligible. On the other hand, we notice that the gyration current coefficient shows a drastic frequency dependence with sign changes in contrast to the magnetic injection current.

\begin{figure}[htbp]
        \centering
            \includegraphics[width=0.90\linewidth,clip]{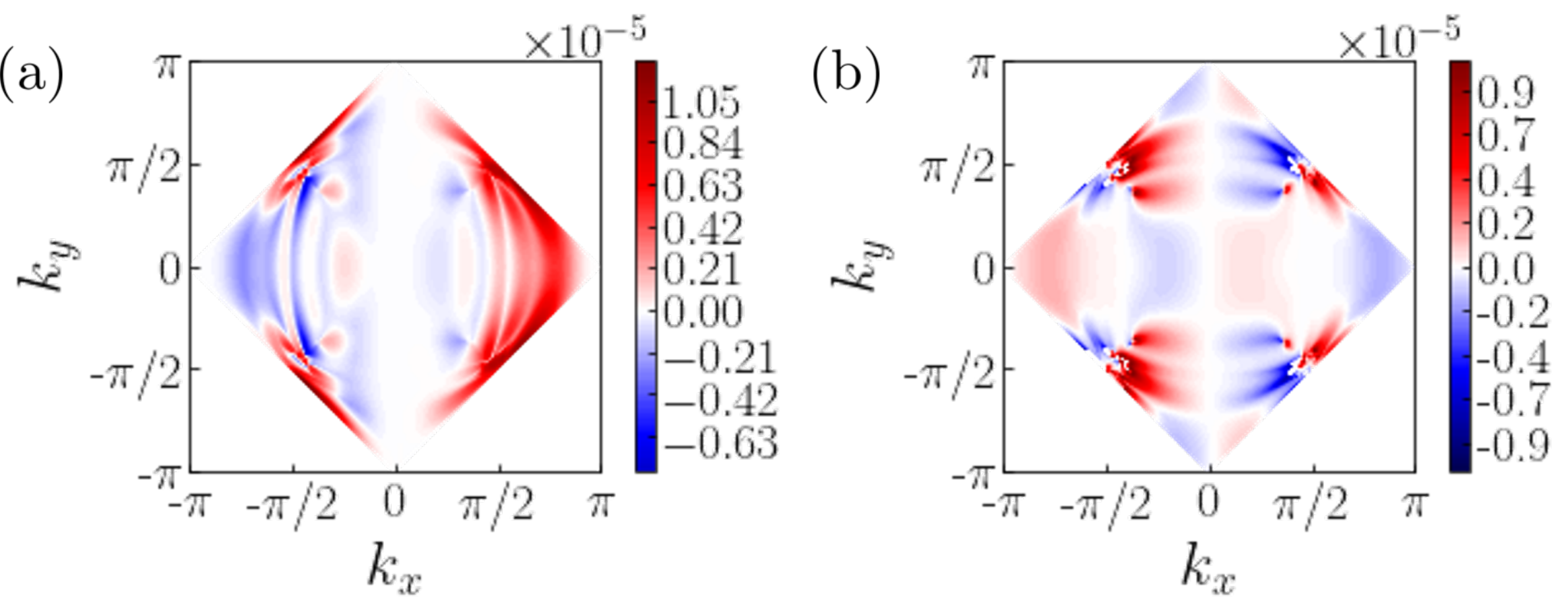}
        \caption{Momentum distribution of (a) the magnetic injection current $\eta_\text{inj}^{x;xx}$ and (b) the gyration current $\kappa_\text{gyro}^{yz}$ in the first Brillouin zone. The parameters are the same as in Fig.~\ref{Fig_attenuation_jdos_plot}. The frequency of light is taken as $ \Omega = \mr{0.61}{eV}$.}
        \label{Fig_photocurrent_kdependence}
        \end{figure}

The frequency dependence may originate from the momentum distribution of the photocurrent responses. With the frequency of light $\Omega \sim \mr{0.5}{eV}$, the optical transitions between $j=1/2$ orbitals intensely happen around the Brillouin zone edge denoted by $k_x\pm k_y =\pm \pi$ due to the generalized van Hove singularity~\cite{HiroshiWatanabe2010,Grosso2013Book}. For the magnetic injection current $\eta^{x;xx}_\text{inj}$, the momentum around the van Hove singularity cooperatively causes the photocurrent generation [Fig.~\ref{Fig_photocurrent_kdependence}(a)]. In contrast, the contribution is partially compensated for the gyration current because of the dipolar-shaped distribution around the van Hove singularity 
[Fig.~\ref{Fig_photocurrent_kdependence}(b)]. Therefore, the gyration current coefficient obtained by summing up in the Brillouin zone shows sign reversal because a slight change in the frequency of lights may deform the dipolar-shaped distribution. 

\begin{figure}[htbp]
        \centering
        \includegraphics[width=0.75\linewidth,clip]{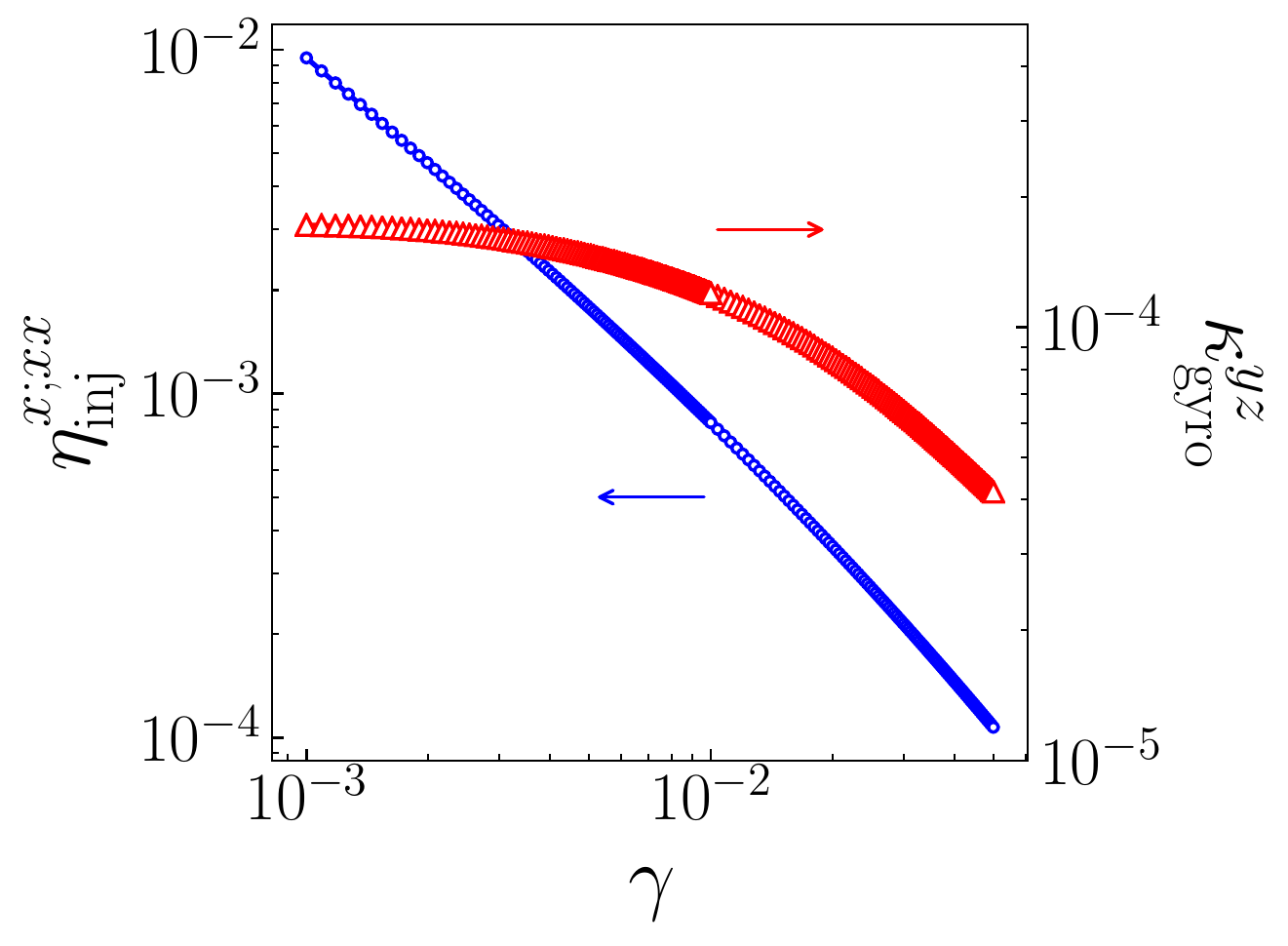}
        \caption{Log-log plot of the scattering rate dependence of the magnetic injection current $\eta_\text{inj}^{x;xx}$ A$\cdot$V$^{-2}$ (blue-colored) and the gyration current $\kappa_\text{gyro}^{yz}$ A$\cdot$V$^{-2}$ (red-colored). The unit of the transverse axis is electron volt (eV).}
        \label{Fig_photocurrent_scatter_dependence}
        \end{figure}

Finally, let us look into the dependence on the scattering rate $\gamma$. The magnetic injection current strongly depends on the scattering rate as seen in Eq.~\eqref{magnetic_injection_current_bare_with_scattering}, and thus, it is suppressed in proportional to $\gamma^{-1}$.
 On the other hand, the gyration current formula [Eq.~\eqref{gyration_current_formula}] does not include any scattering rate other than a smearing factor represented by the Lorentian function. Therefore, the scattering rate dependence shows the different behaviors according to the mechanism of the photocurrent creation (Fig.~\ref{Fig_photocurrent_scatter_dependence}).
With a small scattering rate ($\gamma \sim \mr{0.001}{eV}$),
 the magnetic injection current is much larger than the gyration current, while they are comparable for a moderate scattering ($\gamma \gtrsim \mr{0.05}{eV}$). The robustness to the scattering of the shift current mechanism is supported by a recent experiment in Ref.~\cite{Hatada2020} where the disorder concentration does not influence the shift current response which is the nonmagnetic counterpart of the gyration current. It is noteworthy that for a realistic scattering effect as much as $\gamma \lesssim \mr{0.1}{eV}$, the photocurrent responses in the \PT{} symmetric current-ordered \sro{} are comparable to that in a prototypical semiconductor GaAs, which is theoretically estimated as $\eta^{x;yz} \sim \mr{10}{ \text{\textmu}A \cdot V^{-2}}$~\cite{Nastos2006,Ibanez2018}.

\section{Discussion and Summary}
\label{Sec_discussion_summary}

\begin{figure*}[htbp]
    \centering
        \includegraphics[width=0.95\linewidth,clip]{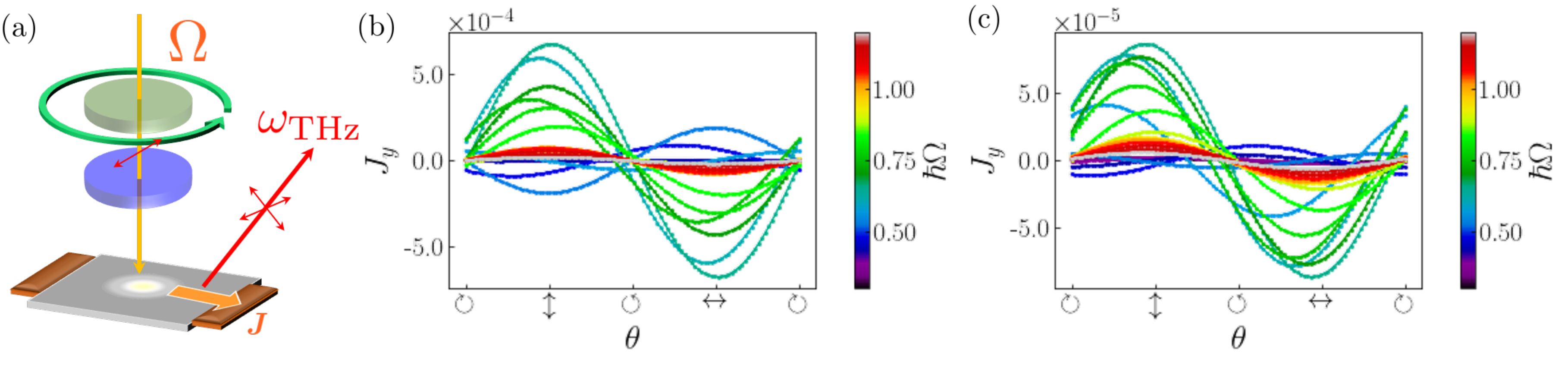}
    \caption{(a) Sketch of an experimental setup for the detection of the photocurrent response. The linearly-polarized light with the frequency $\Omega$ is introduced in the normal incident. The polarization state of the incident light is initialized by the polarizer (green-colored disk) and changed through the quarter-wave plate (blue-colored disk). Created photocurrent $\bm{J}$ flows to the brown-colored contact while it also produces a terahertz-wave radiation denoted by $\omega_\text{THz}$. (b,c) Plots for the polarization state dependence of the photocurrent $J_y$. The symbols `$\updownarrow$' and `$\leftrightarrow$' represent the light linearly-polarized along the $[110]$ and $[-110]$-axes, whereas `$\circlearrowleft$' and `$\circlearrowright$' denote the left-handed and right-handed circularly-polarized states. The color of the plots parametrizes the frequency of irradiating lights. The parameters are  $T=\mr{0.001}{eV}$ and $\gamma=\mr{0.01}{eV}$ ($\gamma=\mr{0.05}{eV}$) in Fig.~\ref{Fig_experimental_setup}(b) [Fig.~\ref{Fig_experimental_setup}(c)]. }
    \label{Fig_experimental_setup}
    \end{figure*}

The calculated photocurrent response comparable to that of typical noncentrosymmetric materials implies the feasibility of the symmetry clarification of \sro{} via the photocurrent detection. By referring to the symmetry analysis in Table~\ref{Table_photocurrent_sro_bond_order_corespondence}, the symmetry of the anapole order may be unveiled. Here we consider an experimental setup for detection of the photocurrent response which has been implemented in prior experiments [see Fig.~\ref{Fig_experimental_setup}(a)]. By changing the polarization state of the irradiating light through the rotation of polarizer or quarter-wave plate, we can measure the photocurrent responses induced by the linearly-polarized and circularly-polarized lights.

The photo-induced electric current is detected through the contact conductivity measurement or by measuring the terahertz lights radiating from flowing photocurrent. In particular, the terahertz-light detection of the photocurrent is less sensitive to photo-thermal effects~\cite{Kastl2015-rj,Takeno2018-gy} and is conveniently distinguished from other contributions such as the spin-galvanic effect~\cite{Kastl2015-rj,Sotome2019,Sirica2019Ultrafast,Gao2020-du}. 

For instance, by rotating the polarizer, the photo-electric field is varied as
		\begin{align}
        \bm{E} = \frac{|\bm{E}_0 |}{\sqrt{2}}\left[  \left(  \cos{\theta} +\sin{\theta}  \right) \hat{e}_1+ i \left(  -\cos{\theta} +\sin{\theta}  \right) \hat{e}_2 \right],
        \end{align}
with the parameter $\theta \in [ 0, \pi ]$ and the unit vectors $\hat{e}_1 \parallel [110]$ and $\hat{e}_2 \parallel [-110]$. Note that the quarter-wave-plate is prepared to modulate the relative phase between linearly-polarized lights along $\hat{e}_i~(i=1,2)$. The left-handed (right-handed) circularly-polarized light corresponds to $\theta= \pi/2$ ($\theta= 0,\pi$), whereas the linearly-polarized light is realized by $\theta = \pi/4,3\pi/4$. Accordingly, the total photocurrent along the $[010]$-direction is evaluated as
		\begin{align}
        J^y 
            &= |\bm{E}_0|^2 \left[  \left(  \sin ^2 \theta - \cos ^2 \theta  \right) \kappa^{yz} + 2 \sin \theta \cos \theta \, \eta^{y;xy} \right],\\
            &= |\bm{E}_0|^2 \left[  -\cos 2 \theta \, \kappa^{yz} + \sin 2 \theta \, \eta^{y;xy} \right],
        \end{align}
in which we take into account the components allowed in the system hosting the $E_u^\pm \Braket{x}$ order (see Table~\ref{Table_photocurrent_sro_bond_order_corespondence}). Based on the model study of \sro{} in Sec.~\ref{Sec_sro_photocurrent}, the dependencies of the photocurrent on the polarization state and the frequency of incident lights are plotted in Figs.~\ref{Fig_experimental_setup}(b) and \ref{Fig_experimental_setup}(c). In Fig.~\ref{Fig_experimental_setup}(b), because the system with a small scattering rate $\gamma = \mr{0.01}{eV}$ is assumed, the photocurrent response is dominantly induced by the linearly-polarized light, and the polarization dependence shows the sine curve. On the other hand, with a moderate scattering rate ($\gamma = \mr{0.05}{eV}$), the magnetic injection current response is suppressed more strongly than the gyration current. Thus, we observe a non-negligible cosine component in the polarization dependence [Fig.~\ref{Fig_experimental_setup}(c)].

Similarly, it is also possible to distinguish the photocurrent response induced by either of the linearly-polarized or circularly-polarized light in another experimental setup.
Here we suppose to match the optical axis of the quarter-wave-plate with the [010]-direction and rotate the polarizer. Accordingly, the polarization state is changed as
        \begin{align}
        \bm{E} = \frac{|\bm{E}_0 |}{\sqrt{2}}
            =\left[  \cos{\theta} \, \hat{x}+ i \sin{\theta}\, \hat{y} \right],
        \end{align}
With this parameterization, the photocurrent along the [010]-direction is given by the circularly-polarized component of light owing to $\kappa^{yz}$, whereas the [100]-photocurrent is determined by the linearly-polarized component.

We can perform the parallel symmetry analysis for the $E_u^\pm \Braket{d}$ current-ordered state. The polarization angle dependence differs between the $E_u^\pm \Braket{x}$ and $E_u^\pm \Braket{d}$ by $\pi/4$ in the $xy$-plane. Thus, the two possible configurations can be distinguished by the photocurrent measurement. For instance, the circularly-polarized light along the $z$-direction produces the photocurrent along the direction perpendicular to the anapole order. Therefore, the direction of the anapole order can be identified by measuring the circularly-photo-induced current in the \sro{} single crystal. For general consideration, let us consider the ferro-anapole order pointed in an arbitrary direction.
We obtain the circular-polarized-light-induced photocurrent response when the incident light, induced photocurrent, and anapole moment are orthogonal to each other.

Our calculation applies to the sufficiently low-temperature regime where the molecular field approximation of the anapole order is reasonable. On the other hand, the Fermi surface effect on the photocurrent response may play a non-negligible role when we work on the doped \sro{} and the systems at higher temperatures~\cite{Xu2020,Cao2016,Louat2018}. Further experimental and theoretical efforts on the nonlinear optical responses in the parent and doped \sro{} may give an implication of the similarity between the iridate and cuprate~\cite{Mitchell2015}.

To summarize, we investigated the photocurrent response arising from a parity-time symmetric anapole order. Motivated by the recent experiment~\cite{Murayama2020}, we performed the symmetry analysis of the possible anapole-ordered states in \sro{} and clarified the photocurrent responses. Making use of the microscopic multi-orbital model of \sro{}, we demonstrated the photocurrent arising from the simultaneous breaking of the \Pa{} and \T{} symmetries. Two types of the photocurrent, magnetic injection current and gyration current, are distinguished by their dependencies on the scattering rate of electrons and frequency and polarization of irradiating lights. In particular, the gyration current is robust to scatterings, consistent with its formula [Eq.~\eqref{gyration_current_formula}]. Interestingly, the calculated magnitude of the photocurrent is comparable to a prototypical value in GaAs. Therefore, the photocurrent response is potentially useful to detect the anapole order in \sro{}.

Although we have not focused on cuprates, the photocurrent response can appear when an anapole (loop-current) order indeed occurs in cuprate superconductors. Thus, the photocurrent response may play a key role in detecting exotic parity-violating phases of matter.

\section*{Acknowledgments}

The authors are grateful to J.~Ishizuka, S.~Kasahara, Y.~Matsuda, and H.~Murayama for fruitful discussions.
This work was supported by JSPS KAKENHI (No. JP18H05227, No. JP18H01178, and No. 20H05159) and SPIRITS
2020 of Kyoto University. H.W. acknowledges support as a JSPS research fellow and supported by JSPS KAKENHI (Grant No.~18J23115).

\appendix

\section{Magnetic injection current in Iridate}\label{Appendix_MIC}
\begin{figure}[htbp]
            \centering
                \includegraphics[width=0.95\linewidth,clip]{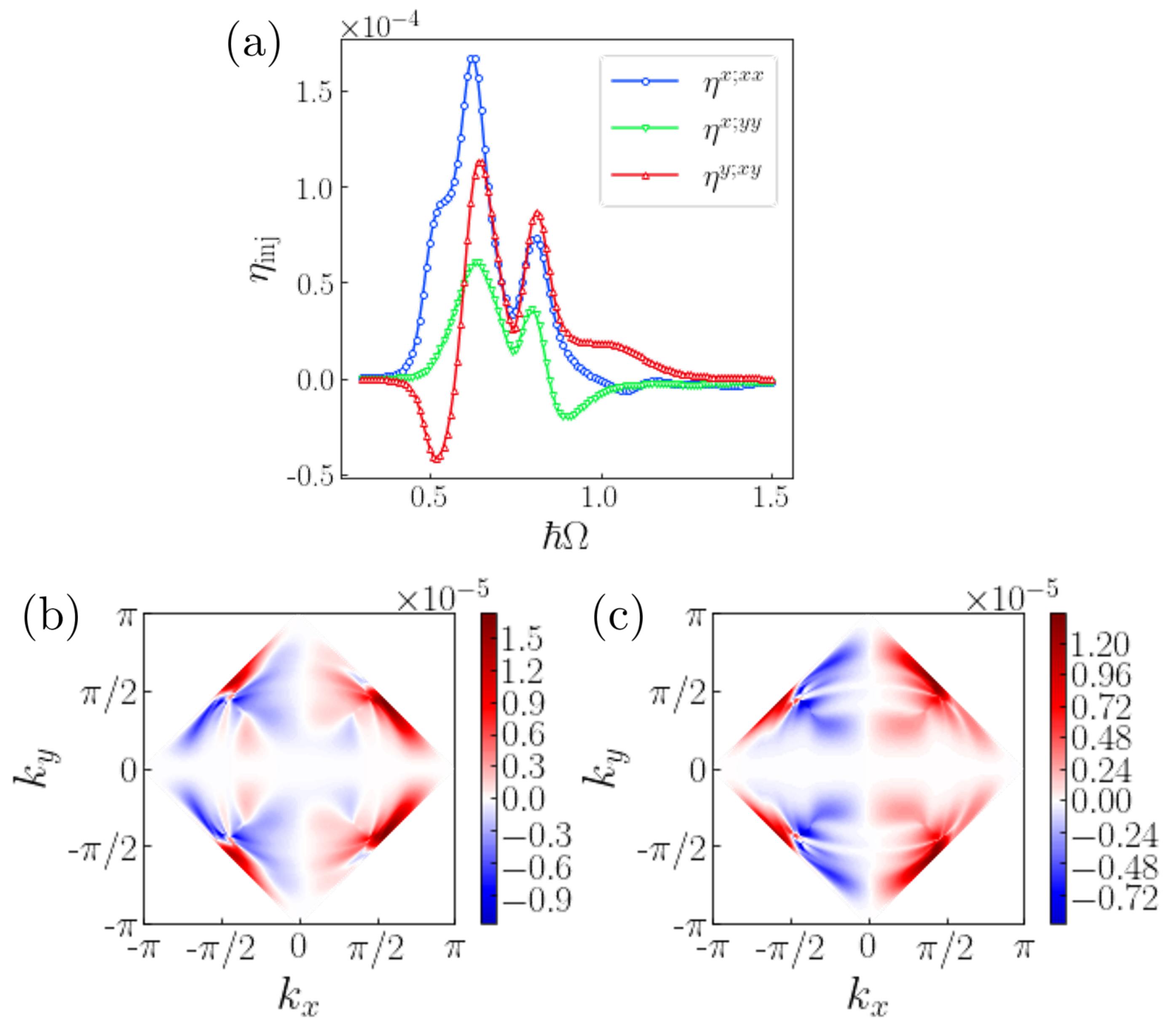}
            \caption{(a) Frequency dependence of the magnetic injection current $\eta^{\mu;\nu\lambda}_\text{inj}$ A$\cdot$V$^{-2}$. All the components allowed by symmetry are plotted;
            $\eta^{x;xx}_\text{inj}$ (blue-colored), $\eta^{x;yy}_\text{inj}$ (red-colored), and $\eta^{y;xy}_\text{inj}$ (green-colored). All the other components are zero. (b,c) Momentum distribution of the magnetic injection currents, (b) $\eta_\text{inj}^{x;yy}$ and (c) $\eta_\text{inj}^{y;xy}$ for $\Omega = \mr{0.61}{eV}$.
            The parameters are the same as in Fig.~\ref{Fig_attenuation_jdos_plot}. 
            }
            \label{Fig_injection_allcomponents_omegadep}
            \end{figure}

In Sec.~\ref{Sec_sro_photocurrent}, we show only one component of the magnetic injection current, $\eta^{x;xx}_\text{inj}$. The other allowed components are 
		\begin{equation}
            \eta^{x;yy}_\text{inj}, \,\,\, \eta^{y;xy}_\text{inj},
        \end{equation}
which respect the $E_u \Braket{x}$ symmetry in Table~\ref{Table_photocurrent_sro_bond_order_corespondence}. The numerical results of these components, $\eta^{x;yy}_\text{inj}$ and $\eta^{y;xy}_\text{inj}$,  are shown in Fig.~\ref{Fig_injection_allcomponents_omegadep}(a). The frequency dependence almost agrees with that of $\eta^{x;xx}_\text{inj}$ in the low-frequency regime $\Omega < \mr{1.0}{eV}$ where an intense optical transition between $j=1/2$ orbitals occurs. The momentum distribution of the components $\eta^{x;yy}_\text{inj},\eta^{y;xy}_\text{inj}$ are plotted in Figs.~\ref{Fig_injection_allcomponents_omegadep}(b) and~\ref{Fig_injection_allcomponents_omegadep}(c). The distributions near the $k_x+k_y=\pi$ and $k_x-k_y=\pi$ lines resemble that for $\eta^{x;xx}_\text{inj}$ [Fig.~\ref{Fig_photocurrent_kdependence}(a)]. On the other hand, we notice a slight difference in the momentum distributions along the $k_x-k_y=-\pi$ and $k_x+k_y=-\pi$ lines between Fig.~\ref{Fig_photocurrent_kdependence}(a) and Figs.~\ref{Fig_injection_allcomponents_omegadep}(b,c). The partial compensation may lead to a sign change of the magnetic injection current response such as what is observed at $\Omega \sim \mr{0.55}{eV}$ for $\eta^{y;xy}_\text{inj}$.

\begin{thebibliography}{65}%
\makeatletter
\providecommand \@ifxundefined [1]{%
 \@ifx{#1\undefined}
}%
\providecommand \@ifnum [1]{%
 \ifnum #1\expandafter \@firstoftwo
 \else \expandafter \@secondoftwo
 \fi
}%
\providecommand \@ifx [1]{%
 \ifx #1\expandafter \@firstoftwo
 \else \expandafter \@secondoftwo
 \fi
}%
\providecommand \natexlab [1]{#1}%
\providecommand \enquote  [1]{``#1''}%
\providecommand \bibnamefont  [1]{#1}%
\providecommand \bibfnamefont [1]{#1}%
\providecommand \citenamefont [1]{#1}%
\providecommand \href@noop [0]{\@secondoftwo}%
\providecommand \href [0]{\begingroup \@sanitize@url \@href}%
\providecommand \@href[1]{\@@startlink{#1}\@@href}%
\providecommand \@@href[1]{\endgroup#1\@@endlink}%
\providecommand \@sanitize@url [0]{\catcode `\\12\catcode `\$12\catcode
  `\&12\catcode `\#12\catcode `\^12\catcode `\_12\catcode `\%12\relax}%
\providecommand \@@startlink[1]{}%
\providecommand \@@endlink[0]{}%
\providecommand \url  [0]{\begingroup\@sanitize@url \@url }%
\providecommand \@url [1]{\endgroup\@href {#1}{\urlprefix }}%
\providecommand \urlprefix  [0]{URL }%
\providecommand \Eprint [0]{\href }%
\providecommand \doibase [0]{http://dx.doi.org/}%
\providecommand \selectlanguage [0]{\@gobble}%
\providecommand \bibinfo  [0]{\@secondoftwo}%
\providecommand \bibfield  [0]{\@secondoftwo}%
\providecommand \translation [1]{[#1]}%
\providecommand \BibitemOpen [0]{}%
\providecommand \bibitemStop [0]{}%
\providecommand \bibitemNoStop [0]{.\EOS\space}%
\providecommand \EOS [0]{\spacefactor3000\relax}%
\providecommand \BibitemShut  [1]{\csname bibitem#1\endcsname}%
\let\auto@bib@innerbib\@empty
\bibitem [{\citenamefont {Sturman}\ and\ \citenamefont
  {Fridkin}(1992)}]{Sturman1992Book}%
  \BibitemOpen
  \bibfield  {author} {\bibinfo {author} {\bibfnamefont {B.~I.}\ \bibnamefont
  {Sturman}}\ and\ \bibinfo {author} {\bibfnamefont {V.~M.}\ \bibnamefont
  {Fridkin}},\ }\href
  {https://www.crcpress.com/Photovoltaic-and-Photo-refractive-Effects-in-Noncentrosymmetric-Materials/Sturman-Fridkin/p/book/9782881244988}
  {\emph {\bibinfo {title} {Photovoltaic and Photo-refractive Effects in
  Noncentrosymmetric Materials}}}\ (\bibinfo  {publisher} {Gordon and Breach},\
  \bibinfo {address} {New York},\ \bibinfo {year} {1992})\BibitemShut {NoStop}%
\bibitem [{\citenamefont {Koch}\ \emph {et~al.}(1976)\citenamefont {Koch},
  \citenamefont {Munser}, \citenamefont {Ruppel},\ and\ \citenamefont
  {W\"urfel}}]{Koch1976BTO}%
  \BibitemOpen
  \bibfield  {author} {\bibinfo {author} {\bibfnamefont {W.~T.~H.}\
  \bibnamefont {Koch}}, \bibinfo {author} {\bibfnamefont {R.}~\bibnamefont
  {Munser}}, \bibinfo {author} {\bibfnamefont {W.}~\bibnamefont {Ruppel}}, \
  and\ \bibinfo {author} {\bibfnamefont {P.}~\bibnamefont {W\"urfel}},\ }\href
  {\doibase 10.1080/00150197608236596} {\bibfield  {journal} {\bibinfo
  {journal} {Ferroelectrics}\ }\textbf {\bibinfo {volume} {13}},\ \bibinfo
  {pages} {305} (\bibinfo {year} {1976})}\BibitemShut {NoStop}%
\bibitem [{\citenamefont {Liu}\ \emph {et~al.}(2020)\citenamefont {Liu},
  \citenamefont {Xia}, \citenamefont {Xiao}, \citenamefont {Garc{\'\i}a~de
  Abajo},\ and\ \citenamefont {Sun}}]{Liu2020SemimetalReview}%
  \BibitemOpen
  \bibfield  {author} {\bibinfo {author} {\bibfnamefont {J.}~\bibnamefont
  {Liu}}, \bibinfo {author} {\bibfnamefont {F.}~\bibnamefont {Xia}}, \bibinfo
  {author} {\bibfnamefont {D.}~\bibnamefont {Xiao}}, \bibinfo {author}
  {\bibfnamefont {F.~J.}\ \bibnamefont {Garc{\'\i}a~de Abajo}}, \ and\ \bibinfo
  {author} {\bibfnamefont {D.}~\bibnamefont {Sun}},\ }\href
  {http://dx.doi.org/10.1038/s41563-020-0715-7} {\bibfield  {journal} {\bibinfo
   {journal} {Nat. Mater.}\ }\textbf {\bibinfo {volume} {19}},\ \bibinfo
  {pages} {830} (\bibinfo {year} {2020})}\BibitemShut {NoStop}%
\bibitem [{\citenamefont {Sipe}\ and\ \citenamefont
  {Shkrebtii}(2000)}]{Sipe2000secondorder}%
  \BibitemOpen
  \bibfield  {author} {\bibinfo {author} {\bibfnamefont {J.~E.}\ \bibnamefont
  {Sipe}}\ and\ \bibinfo {author} {\bibfnamefont {A.~I.}\ \bibnamefont
  {Shkrebtii}},\ }\href {\doibase 10.1103/PhysRevB.61.5337} {\bibfield
  {journal} {\bibinfo  {journal} {Phys. Rev. B}\ }\textbf {\bibinfo {volume}
  {61}},\ \bibinfo {pages} {5337} (\bibinfo {year} {2000})}\BibitemShut
  {NoStop}%
\bibitem [{\citenamefont {Ventura}\ \emph {et~al.}(2017)\citenamefont
  {Ventura}, \citenamefont {Passos}, \citenamefont {Lopes~dos Santos},
  \citenamefont {Viana Parente~Lopes},\ and\ \citenamefont
  {Peres}}]{Ventura2017}%
  \BibitemOpen
  \bibfield  {author} {\bibinfo {author} {\bibfnamefont {G.~B.}\ \bibnamefont
  {Ventura}}, \bibinfo {author} {\bibfnamefont {D.~J.}\ \bibnamefont {Passos}},
  \bibinfo {author} {\bibfnamefont {J.~M.~B.}\ \bibnamefont {Lopes~dos
  Santos}}, \bibinfo {author} {\bibfnamefont {J.~M.}\ \bibnamefont {Viana
  Parente~Lopes}}, \ and\ \bibinfo {author} {\bibfnamefont {N.~M.~R.}\
  \bibnamefont {Peres}},\ }\href {\doibase 10.1103/PhysRevB.96.035431}
  {\bibfield  {journal} {\bibinfo  {journal} {Phys. Rev. B}\ }\textbf {\bibinfo
  {volume} {96}},\ \bibinfo {pages} {035431} (\bibinfo {year}
  {2017})}\BibitemShut {NoStop}%
\bibitem [{\citenamefont {Passos}\ \emph {et~al.}(2018)\citenamefont {Passos},
  \citenamefont {Ventura}, \citenamefont {Lopes}, \citenamefont {Santos},\ and\
  \citenamefont {Peres}}]{Passos2018}%
  \BibitemOpen
  \bibfield  {author} {\bibinfo {author} {\bibfnamefont {D.~J.}\ \bibnamefont
  {Passos}}, \bibinfo {author} {\bibfnamefont {G.~B.}\ \bibnamefont {Ventura}},
  \bibinfo {author} {\bibfnamefont {J.~M. V.~P.}\ \bibnamefont {Lopes}},
  \bibinfo {author} {\bibfnamefont {J.~M. B. L.~d.}\ \bibnamefont {Santos}}, \
  and\ \bibinfo {author} {\bibfnamefont {N.~M.~R.}\ \bibnamefont {Peres}},\
  }\href {\doibase 10.1103/PhysRevB.97.235446} {\bibfield  {journal} {\bibinfo
  {journal} {Phys. Rev. B}\ }\textbf {\bibinfo {volume} {97}},\ \bibinfo
  {pages} {235446} (\bibinfo {year} {2018})}\BibitemShut {NoStop}%
\bibitem [{\citenamefont {Parker}\ \emph {et~al.}(2019)\citenamefont {Parker},
  \citenamefont {Morimoto}, \citenamefont {Orenstein},\ and\ \citenamefont
  {Moore}}]{Parker2019}%
  \BibitemOpen
  \bibfield  {author} {\bibinfo {author} {\bibfnamefont {D.~E.}\ \bibnamefont
  {Parker}}, \bibinfo {author} {\bibfnamefont {T.}~\bibnamefont {Morimoto}},
  \bibinfo {author} {\bibfnamefont {J.}~\bibnamefont {Orenstein}}, \ and\
  \bibinfo {author} {\bibfnamefont {J.~E.}\ \bibnamefont {Moore}},\ }\href
  {\doibase 10.1103/PhysRevB.99.045121} {\bibfield  {journal} {\bibinfo
  {journal} {Phys. Rev. B}\ }\textbf {\bibinfo {volume} {99}},\ \bibinfo
  {pages} {045121} (\bibinfo {year} {2019})}\BibitemShut {NoStop}%
\bibitem [{\citenamefont {Jo{\~a}o}\ and\ \citenamefont {Viana
  Parente~Lopes}(2020)}]{Joao2020-mk}%
  \BibitemOpen
  \bibfield  {author} {\bibinfo {author} {\bibfnamefont {S.~M.}\ \bibnamefont
  {Jo{\~a}o}}\ and\ \bibinfo {author} {\bibfnamefont {J.~M.}\ \bibnamefont
  {Viana Parente~Lopes}},\ }\href {http://dx.doi.org/10.1088/1361-648X/ab59ec}
  {\bibfield  {journal} {\bibinfo  {journal} {J. Phys. Condens. Matter}\
  }\textbf {\bibinfo {volume} {32}},\ \bibinfo {pages} {125901} (\bibinfo
  {year} {2020})}\BibitemShut {NoStop}%
\bibitem [{\citenamefont {Michishita}\ and\ \citenamefont
  {Peters}(2021)}]{Michishita2020}%
  \BibitemOpen
  \bibfield  {author} {\bibinfo {author} {\bibfnamefont {Y.}~\bibnamefont
  {Michishita}}\ and\ \bibinfo {author} {\bibfnamefont {R.}~\bibnamefont
  {Peters}},\ }\href {\doibase 10.1103/PhysRevB.103.195133} {\bibfield
  {journal} {\bibinfo  {journal} {Phys. Rev. B}\ }\textbf {\bibinfo {volume}
  {103}},\ \bibinfo {pages} {195133} (\bibinfo {year} {2021})}\BibitemShut
  {NoStop}%
\bibitem [{\citenamefont {Young}\ and\ \citenamefont
  {Rappe}(2012)}]{YoungRappe2012_FirstPrincipleBTO}%
  \BibitemOpen
  \bibfield  {author} {\bibinfo {author} {\bibfnamefont {S.~M.}\ \bibnamefont
  {Young}}\ and\ \bibinfo {author} {\bibfnamefont {A.~M.}\ \bibnamefont
  {Rappe}},\ }\href {\doibase 10.1103/PhysRevLett.109.116601} {\bibfield
  {journal} {\bibinfo  {journal} {Phys. Rev. Lett.}\ }\textbf {\bibinfo
  {volume} {109}},\ \bibinfo {pages} {116601} (\bibinfo {year}
  {2012})}\BibitemShut {NoStop}%
\bibitem [{\citenamefont {Young}\ \emph {et~al.}(2012)\citenamefont {Young},
  \citenamefont {Zheng},\ and\ \citenamefont {Rappe}}]{YoungRappe2012BiFeO3}%
  \BibitemOpen
  \bibfield  {author} {\bibinfo {author} {\bibfnamefont {S.~M.}\ \bibnamefont
  {Young}}, \bibinfo {author} {\bibfnamefont {F.}~\bibnamefont {Zheng}}, \ and\
  \bibinfo {author} {\bibfnamefont {A.~M.}\ \bibnamefont {Rappe}},\ }\href
  {\doibase 10.1103/PhysRevLett.109.236601} {\bibfield  {journal} {\bibinfo
  {journal} {Phys. Rev. Lett.}\ }\textbf {\bibinfo {volume} {109}},\ \bibinfo
  {pages} {236601} (\bibinfo {year} {2012})}\BibitemShut {NoStop}%
\bibitem [{\citenamefont {Zhang}\ \emph {et~al.}(2019)\citenamefont {Zhang},
  \citenamefont {Holder}, \citenamefont {Ishizuka}, \citenamefont {de~Juan},
  \citenamefont {Nagaosa}, \citenamefont {Felser},\ and\ \citenamefont
  {Yan}}]{Zhang2019switchable}%
  \BibitemOpen
  \bibfield  {author} {\bibinfo {author} {\bibfnamefont {Y.}~\bibnamefont
  {Zhang}}, \bibinfo {author} {\bibfnamefont {T.}~\bibnamefont {Holder}},
  \bibinfo {author} {\bibfnamefont {H.}~\bibnamefont {Ishizuka}}, \bibinfo
  {author} {\bibfnamefont {F.}~\bibnamefont {de~Juan}}, \bibinfo {author}
  {\bibfnamefont {N.}~\bibnamefont {Nagaosa}}, \bibinfo {author} {\bibfnamefont
  {C.}~\bibnamefont {Felser}}, \ and\ \bibinfo {author} {\bibfnamefont
  {B.}~\bibnamefont {Yan}},\ }\href
  {https://www.nature.com/articles/s41467-019-11832-3} {\bibfield  {journal}
  {\bibinfo  {journal} {Nat. Commun.}\ }\textbf {\bibinfo {volume} {10}},\
  \bibinfo {pages} {1} (\bibinfo {year} {2019})}\BibitemShut {NoStop}%
\bibitem [{\citenamefont {Watanabe}\ and\ \citenamefont
  {Yanase}(2021)}]{watanabe2020chiral}%
  \BibitemOpen
  \bibfield  {author} {\bibinfo {author} {\bibfnamefont {H.}~\bibnamefont
  {Watanabe}}\ and\ \bibinfo {author} {\bibfnamefont {Y.}~\bibnamefont
  {Yanase}},\ }\href {\doibase 10.1103/PhysRevX.11.011001} {\bibfield
  {journal} {\bibinfo  {journal} {Phys. Rev. X}\ }\textbf {\bibinfo {volume}
  {11}},\ \bibinfo {pages} {011001} (\bibinfo {year} {2021})}\BibitemShut
  {NoStop}%
\bibitem [{\citenamefont {Ahn}\ \emph {et~al.}(2020)\citenamefont {Ahn},
  \citenamefont {Guo},\ and\ \citenamefont {Nagaosa}}]{Ahn2020}%
  \BibitemOpen
  \bibfield  {author} {\bibinfo {author} {\bibfnamefont {J.}~\bibnamefont
  {Ahn}}, \bibinfo {author} {\bibfnamefont {G.-Y.}\ \bibnamefont {Guo}}, \ and\
  \bibinfo {author} {\bibfnamefont {N.}~\bibnamefont {Nagaosa}},\ }\href
  {\doibase 10.1103/PhysRevX.10.041041} {\bibfield  {journal} {\bibinfo
  {journal} {Phys. Rev. X}\ }\textbf {\bibinfo {volume} {10}},\ \bibinfo
  {pages} {041041} (\bibinfo {year} {2020})}\BibitemShut {NoStop}%
\bibitem [{\citenamefont {Watanabe}\ and\ \citenamefont
  {Yanase}(2018)}]{Watanabe2018grouptheoretical}%
  \BibitemOpen
  \bibfield  {author} {\bibinfo {author} {\bibfnamefont {H.}~\bibnamefont
  {Watanabe}}\ and\ \bibinfo {author} {\bibfnamefont {Y.}~\bibnamefont
  {Yanase}},\ }\href {\doibase 10.1103/PhysRevB.98.245129} {\bibfield
  {journal} {\bibinfo  {journal} {Phys. Rev. B}\ }\textbf {\bibinfo {volume}
  {98}},\ \bibinfo {pages} {245129} (\bibinfo {year} {2018})}\BibitemShut
  {NoStop}%
\bibitem [{\citenamefont {Hayami}\ \emph {et~al.}(2018)\citenamefont {Hayami},
  \citenamefont {Yatsushiro}, \citenamefont {Yanagi},\ and\ \citenamefont
  {Kusunose}}]{Hayami2018Classification}%
  \BibitemOpen
  \bibfield  {author} {\bibinfo {author} {\bibfnamefont {S.}~\bibnamefont
  {Hayami}}, \bibinfo {author} {\bibfnamefont {M.}~\bibnamefont {Yatsushiro}},
  \bibinfo {author} {\bibfnamefont {Y.}~\bibnamefont {Yanagi}}, \ and\ \bibinfo
  {author} {\bibfnamefont {H.}~\bibnamefont {Kusunose}},\ }\href {\doibase
  10.1103/PhysRevB.98.165110} {\bibfield  {journal} {\bibinfo  {journal} {Phys.
  Rev. B}\ }\textbf {\bibinfo {volume} {98}},\ \bibinfo {pages} {165110}
  (\bibinfo {year} {2018})}\BibitemShut {NoStop}%
\bibitem [{\citenamefont {Maruyama}\ \emph {et~al.}(2012)\citenamefont
  {Maruyama}, \citenamefont {Sigrist},\ and\ \citenamefont
  {Yanase}}]{Maruyama2012}%
  \BibitemOpen
  \bibfield  {author} {\bibinfo {author} {\bibfnamefont {D.}~\bibnamefont
  {Maruyama}}, \bibinfo {author} {\bibfnamefont {M.}~\bibnamefont {Sigrist}}, \
  and\ \bibinfo {author} {\bibfnamefont {Y.}~\bibnamefont {Yanase}},\ }\href
  {\doibase 10.1143/JPSJ.81.034702} {\bibfield  {journal} {\bibinfo  {journal}
  {J. Phys. Soc. Japan}\ }\textbf {\bibinfo {volume} {81}},\ \bibinfo {pages}
  {034702} (\bibinfo {year} {2012})}\BibitemShut {NoStop}%
\bibitem [{\citenamefont {Yanase}(2014)}]{Yanase2014zigzag}%
  \BibitemOpen
  \bibfield  {author} {\bibinfo {author} {\bibfnamefont {Y.}~\bibnamefont
  {Yanase}},\ }\href {\doibase 10.7566/JPSJ.83.014703} {\bibfield  {journal}
  {\bibinfo  {journal} {J. Phys. Soc. Japan}\ }\textbf {\bibinfo {volume}
  {83}},\ \bibinfo {pages} {014703} (\bibinfo {year} {2014})}\BibitemShut
  {NoStop}%
\bibitem [{\citenamefont {Hayami}\ \emph {et~al.}(2014)\citenamefont {Hayami},
  \citenamefont {Kusunose},\ and\ \citenamefont {Motome}}]{Hayami2014}%
  \BibitemOpen
  \bibfield  {author} {\bibinfo {author} {\bibfnamefont {S.}~\bibnamefont
  {Hayami}}, \bibinfo {author} {\bibfnamefont {H.}~\bibnamefont {Kusunose}}, \
  and\ \bibinfo {author} {\bibfnamefont {Y.}~\bibnamefont {Motome}},\ }\href
  {\doibase 10.1103/PhysRevB.90.024432} {\bibfield  {journal} {\bibinfo
  {journal} {Phys. Rev. B}\ }\textbf {\bibinfo {volume} {90}},\ \bibinfo
  {pages} {024432} (\bibinfo {year} {2014})}\BibitemShut {NoStop}%
\bibitem [{\citenamefont {Holder}\ \emph {et~al.}(2020)\citenamefont {Holder},
  \citenamefont {Kaplan},\ and\ \citenamefont {Yan}}]{Holder2020consequences}%
  \BibitemOpen
  \bibfield  {author} {\bibinfo {author} {\bibfnamefont {T.}~\bibnamefont
  {Holder}}, \bibinfo {author} {\bibfnamefont {D.}~\bibnamefont {Kaplan}}, \
  and\ \bibinfo {author} {\bibfnamefont {B.}~\bibnamefont {Yan}},\ }\href
  {https://link.aps.org/doi/10.1103/PhysRevResearch.2.033100} {\bibfield
  {journal} {\bibinfo  {journal} {Phys. Rev. Research}\ }\textbf {\bibinfo
  {volume} {2}},\ \bibinfo {pages} {033100} (\bibinfo {year}
  {2020})}\BibitemShut {NoStop}%
\bibitem [{\citenamefont {Fei}\ \emph {et~al.}(2020)\citenamefont {Fei},
  \citenamefont {Song},\ and\ \citenamefont {Yang}}]{fei2020giant}%
  \BibitemOpen
  \bibfield  {author} {\bibinfo {author} {\bibfnamefont {R.}~\bibnamefont
  {Fei}}, \bibinfo {author} {\bibfnamefont {W.}~\bibnamefont {Song}}, \ and\
  \bibinfo {author} {\bibfnamefont {L.}~\bibnamefont {Yang}},\ }\href
  {https://link.aps.org/doi/10.1103/PhysRevB.102.035440} {\bibfield  {journal}
  {\bibinfo  {journal} {Phys. Rev. B}\ }\textbf {\bibinfo {volume} {102}},\
  \bibinfo {pages} {035440} (\bibinfo {year} {2020})}\BibitemShut {NoStop}%
\bibitem [{\citenamefont {Varma}(1997)}]{Varma1997}%
  \BibitemOpen
  \bibfield  {author} {\bibinfo {author} {\bibfnamefont {C.~M.}\ \bibnamefont
  {Varma}},\ }\href {\doibase 10.1103/PhysRevB.55.14554} {\bibfield  {journal}
  {\bibinfo  {journal} {Phys. Rev. B}\ }\textbf {\bibinfo {volume} {55}},\
  \bibinfo {pages} {14554} (\bibinfo {year} {1997})}\BibitemShut {NoStop}%
\bibitem [{\citenamefont {Scagnoli}\ \emph {et~al.}(2011)\citenamefont
  {Scagnoli}, \citenamefont {Staub}, \citenamefont {Bodenthin}, \citenamefont
  {de~Souza}, \citenamefont {Garc{\'\i}a-Fern{\'a}ndez}, \citenamefont
  {Garganourakis}, \citenamefont {Boothroyd}, \citenamefont {Prabhakaran},\
  and\ \citenamefont {Lovesey}}]{Scagnoli2011-tg}%
  \BibitemOpen
  \bibfield  {author} {\bibinfo {author} {\bibfnamefont {V.}~\bibnamefont
  {Scagnoli}}, \bibinfo {author} {\bibfnamefont {U.}~\bibnamefont {Staub}},
  \bibinfo {author} {\bibfnamefont {Y.}~\bibnamefont {Bodenthin}}, \bibinfo
  {author} {\bibfnamefont {R.~A.}\ \bibnamefont {de~Souza}}, \bibinfo {author}
  {\bibfnamefont {M.}~\bibnamefont {Garc{\'\i}a-Fern{\'a}ndez}}, \bibinfo
  {author} {\bibfnamefont {M.}~\bibnamefont {Garganourakis}}, \bibinfo {author}
  {\bibfnamefont {A.~T.}\ \bibnamefont {Boothroyd}}, \bibinfo {author}
  {\bibfnamefont {D.}~\bibnamefont {Prabhakaran}}, \ and\ \bibinfo {author}
  {\bibfnamefont {S.~W.}\ \bibnamefont {Lovesey}},\ }\href
  {http://dx.doi.org/10.1126/science.1201061} {\bibfield  {journal} {\bibinfo
  {journal} {Science}\ }\textbf {\bibinfo {volume} {332}},\ \bibinfo {pages}
  {696} (\bibinfo {year} {2011})}\BibitemShut {NoStop}%
\bibitem [{\citenamefont {Fauqu\'e}\ \emph {et~al.}(2006)\citenamefont
  {Fauqu\'e}, \citenamefont {Sidis}, \citenamefont {Hinkov}, \citenamefont
  {Pailh\`es}, \citenamefont {Lin}, \citenamefont {Chaud},\ and\ \citenamefont
  {Bourges}}]{Fauque2006}%
  \BibitemOpen
  \bibfield  {author} {\bibinfo {author} {\bibfnamefont {B.}~\bibnamefont
  {Fauqu\'e}}, \bibinfo {author} {\bibfnamefont {Y.}~\bibnamefont {Sidis}},
  \bibinfo {author} {\bibfnamefont {V.}~\bibnamefont {Hinkov}}, \bibinfo
  {author} {\bibfnamefont {S.}~\bibnamefont {Pailh\`es}}, \bibinfo {author}
  {\bibfnamefont {C.~T.}\ \bibnamefont {Lin}}, \bibinfo {author} {\bibfnamefont
  {X.}~\bibnamefont {Chaud}}, \ and\ \bibinfo {author} {\bibfnamefont
  {P.}~\bibnamefont {Bourges}},\ }\href {\doibase
  10.1103/PhysRevLett.96.197001} {\bibfield  {journal} {\bibinfo  {journal}
  {Phys. Rev. Lett.}\ }\textbf {\bibinfo {volume} {96}},\ \bibinfo {pages}
  {197001} (\bibinfo {year} {2006})}\BibitemShut {NoStop}%
\bibitem [{\citenamefont {Li}\ \emph {et~al.}(2008)\citenamefont {Li},
  \citenamefont {Bal{\'e}dent}, \citenamefont {Barisi{\'c}}, \citenamefont
  {Cho}, \citenamefont {Fauqu{\'e}}, \citenamefont {Sidis}, \citenamefont {Yu},
  \citenamefont {Zhao}, \citenamefont {Bourges},\ and\ \citenamefont
  {Greven}}]{Li2008-hj}%
  \BibitemOpen
  \bibfield  {author} {\bibinfo {author} {\bibfnamefont {Y.}~\bibnamefont
  {Li}}, \bibinfo {author} {\bibfnamefont {V.}~\bibnamefont {Bal{\'e}dent}},
  \bibinfo {author} {\bibfnamefont {N.}~\bibnamefont {Barisi{\'c}}}, \bibinfo
  {author} {\bibfnamefont {Y.}~\bibnamefont {Cho}}, \bibinfo {author}
  {\bibfnamefont {B.}~\bibnamefont {Fauqu{\'e}}}, \bibinfo {author}
  {\bibfnamefont {Y.}~\bibnamefont {Sidis}}, \bibinfo {author} {\bibfnamefont
  {G.}~\bibnamefont {Yu}}, \bibinfo {author} {\bibfnamefont {X.}~\bibnamefont
  {Zhao}}, \bibinfo {author} {\bibfnamefont {P.}~\bibnamefont {Bourges}}, \
  and\ \bibinfo {author} {\bibfnamefont {M.}~\bibnamefont {Greven}},\ }\href
  {http://dx.doi.org/10.1038/nature07251} {\bibfield  {journal} {\bibinfo
  {journal} {Nature}\ }\textbf {\bibinfo {volume} {455}},\ \bibinfo {pages}
  {372} (\bibinfo {year} {2008})}\BibitemShut {NoStop}%
\bibitem [{\citenamefont {Pershoguba}\ \emph {et~al.}(2013)\citenamefont
  {Pershoguba}, \citenamefont {Kechedzhi},\ and\ \citenamefont
  {Yakovenko}}]{Pershoguba2013-jy}%
  \BibitemOpen
  \bibfield  {author} {\bibinfo {author} {\bibfnamefont {S.~S.}\ \bibnamefont
  {Pershoguba}}, \bibinfo {author} {\bibfnamefont {K.}~\bibnamefont
  {Kechedzhi}}, \ and\ \bibinfo {author} {\bibfnamefont {V.~M.}\ \bibnamefont
  {Yakovenko}},\ }\href {http://dx.doi.org/10.1103/PhysRevLett.111.047005}
  {\bibfield  {journal} {\bibinfo  {journal} {Phys. Rev. Lett.}\ }\textbf
  {\bibinfo {volume} {111}},\ \bibinfo {pages} {047005} (\bibinfo {year}
  {2013})}\BibitemShut {NoStop}%
\bibitem [{\citenamefont {Pershoguba}\ \emph {et~al.}(2014)\citenamefont
  {Pershoguba}, \citenamefont {Kechedzhi},\ and\ \citenamefont
  {Yakovenko}}]{Pershoguba2014-dh}%
  \BibitemOpen
  \bibfield  {author} {\bibinfo {author} {\bibfnamefont {S.~S.}\ \bibnamefont
  {Pershoguba}}, \bibinfo {author} {\bibfnamefont {K.}~\bibnamefont
  {Kechedzhi}}, \ and\ \bibinfo {author} {\bibfnamefont {V.~M.}\ \bibnamefont
  {Yakovenko}},\ }\href
  {https://link.aps.org/doi/10.1103/PhysRevLett.113.129901} {\bibfield
  {journal} {\bibinfo  {journal} {Phys. Rev. Lett.}\ }\textbf {\bibinfo
  {volume} {113}},\ \bibinfo {pages} {129901} (\bibinfo {year}
  {2014})}\BibitemShut {NoStop}%
\bibitem [{\citenamefont {Zhao}\ \emph {et~al.}(2017)\citenamefont {Zhao},
  \citenamefont {Belvin}, \citenamefont {Liang}, \citenamefont {Bonn},
  \citenamefont {Hardy}, \citenamefont {Armitage},\ and\ \citenamefont
  {Hsieh}}]{Zhao2017-ka}%
  \BibitemOpen
  \bibfield  {author} {\bibinfo {author} {\bibfnamefont {L.}~\bibnamefont
  {Zhao}}, \bibinfo {author} {\bibfnamefont {C.~A.}\ \bibnamefont {Belvin}},
  \bibinfo {author} {\bibfnamefont {R.}~\bibnamefont {Liang}}, \bibinfo
  {author} {\bibfnamefont {D.~A.}\ \bibnamefont {Bonn}}, \bibinfo {author}
  {\bibfnamefont {W.~N.}\ \bibnamefont {Hardy}}, \bibinfo {author}
  {\bibfnamefont {N.~P.}\ \bibnamefont {Armitage}}, \ and\ \bibinfo {author}
  {\bibfnamefont {D.}~\bibnamefont {Hsieh}},\ }\href
  {https://doi.org/10.1038/nphys3962} {\bibfield  {journal} {\bibinfo
  {journal} {Nat. Phys.}\ }\textbf {\bibinfo {volume} {13}},\ \bibinfo {pages}
  {250} (\bibinfo {year} {2017})}\BibitemShut {NoStop}%
\bibitem [{\citenamefont {Zhao}\ \emph {et~al.}(2016)\citenamefont {Zhao},
  \citenamefont {Torchinsky}, \citenamefont {Chu}, \citenamefont {Ivanov},
  \citenamefont {Lifshitz}, \citenamefont {Flint}, \citenamefont {Qi},
  \citenamefont {Cao},\ and\ \citenamefont {Hsieh}}]{Zhao2016SHG}%
  \BibitemOpen
  \bibfield  {author} {\bibinfo {author} {\bibfnamefont {L.}~\bibnamefont
  {Zhao}}, \bibinfo {author} {\bibfnamefont {D.~H.}\ \bibnamefont
  {Torchinsky}}, \bibinfo {author} {\bibfnamefont {H.}~\bibnamefont {Chu}},
  \bibinfo {author} {\bibfnamefont {V.}~\bibnamefont {Ivanov}}, \bibinfo
  {author} {\bibfnamefont {R.}~\bibnamefont {Lifshitz}}, \bibinfo {author}
  {\bibfnamefont {R.}~\bibnamefont {Flint}}, \bibinfo {author} {\bibfnamefont
  {T.}~\bibnamefont {Qi}}, \bibinfo {author} {\bibfnamefont {G.}~\bibnamefont
  {Cao}}, \ and\ \bibinfo {author} {\bibfnamefont {D.}~\bibnamefont {Hsieh}},\
  }\href {\doibase 10.1038/nphys3517} {\bibfield  {journal} {\bibinfo
  {journal} {Nat. Phys.}\ }\textbf {\bibinfo {volume} {12}},\ \bibinfo {pages}
  {32} (\bibinfo {year} {2016})}\BibitemShut {NoStop}%
\bibitem [{\citenamefont {Jeong}\ \emph {et~al.}(2017)\citenamefont {Jeong},
  \citenamefont {Sidis}, \citenamefont {Louat}, \citenamefont {Brouet},\ and\
  \citenamefont {Bourges}}]{Jeong2017}%
  \BibitemOpen
  \bibfield  {author} {\bibinfo {author} {\bibfnamefont {J.}~\bibnamefont
  {Jeong}}, \bibinfo {author} {\bibfnamefont {Y.}~\bibnamefont {Sidis}},
  \bibinfo {author} {\bibfnamefont {A.}~\bibnamefont {Louat}}, \bibinfo
  {author} {\bibfnamefont {V.}~\bibnamefont {Brouet}}, \ and\ \bibinfo {author}
  {\bibfnamefont {P.}~\bibnamefont {Bourges}},\ }\href
  {http://dx.doi.org/10.1038/ncomms15119} {\bibfield  {journal} {\bibinfo
  {journal} {Nat. Commun.}\ }\textbf {\bibinfo {volume} {8}},\ \bibinfo {pages}
  {15119} (\bibinfo {year} {2017})}\BibitemShut {NoStop}%
\bibitem [{\citenamefont {Murayama}\ \emph {et~al.}(2021)\citenamefont
  {Murayama}, \citenamefont {Ishida}, \citenamefont {Kurihara}, \citenamefont
  {Ono}, \citenamefont {Sato}, \citenamefont {Kasahara}, \citenamefont
  {Watanabe}, \citenamefont {Yanase}, \citenamefont {Cao}, \citenamefont
  {Mizukami}, \citenamefont {Shibauchi}, \citenamefont {Matsuda},\ and\
  \citenamefont {Kasahara}}]{Murayama2020}%
  \BibitemOpen
  \bibfield  {author} {\bibinfo {author} {\bibfnamefont {H.}~\bibnamefont
  {Murayama}}, \bibinfo {author} {\bibfnamefont {K.}~\bibnamefont {Ishida}},
  \bibinfo {author} {\bibfnamefont {R.}~\bibnamefont {Kurihara}}, \bibinfo
  {author} {\bibfnamefont {T.}~\bibnamefont {Ono}}, \bibinfo {author}
  {\bibfnamefont {Y.}~\bibnamefont {Sato}}, \bibinfo {author} {\bibfnamefont
  {Y.}~\bibnamefont {Kasahara}}, \bibinfo {author} {\bibfnamefont
  {H.}~\bibnamefont {Watanabe}}, \bibinfo {author} {\bibfnamefont
  {Y.}~\bibnamefont {Yanase}}, \bibinfo {author} {\bibfnamefont
  {G.}~\bibnamefont {Cao}}, \bibinfo {author} {\bibfnamefont {Y.}~\bibnamefont
  {Mizukami}}, \bibinfo {author} {\bibfnamefont {T.}~\bibnamefont {Shibauchi}},
  \bibinfo {author} {\bibfnamefont {Y.}~\bibnamefont {Matsuda}}, \ and\
  \bibinfo {author} {\bibfnamefont {S.}~\bibnamefont {Kasahara}},\ }\href
  {\doibase 10.1103/PhysRevX.11.011021} {\bibfield  {journal} {\bibinfo
  {journal} {Phys. Rev. X}\ }\textbf {\bibinfo {volume} {11}},\ \bibinfo
  {pages} {011021} (\bibinfo {year} {2021})}\BibitemShut {NoStop}%
\bibitem [{\citenamefont {Van~Aken}\ \emph {et~al.}(2007)\citenamefont
  {Van~Aken}, \citenamefont {Rivera}, \citenamefont {Schmid},\ and\
  \citenamefont {Fiebig}}]{Van_Aken2007-zf}%
  \BibitemOpen
  \bibfield  {author} {\bibinfo {author} {\bibfnamefont {B.~B.}\ \bibnamefont
  {Van~Aken}}, \bibinfo {author} {\bibfnamefont {J.-P.}\ \bibnamefont
  {Rivera}}, \bibinfo {author} {\bibfnamefont {H.}~\bibnamefont {Schmid}}, \
  and\ \bibinfo {author} {\bibfnamefont {M.}~\bibnamefont {Fiebig}},\ }\href
  {http://dx.doi.org/10.1038/nature06139} {\bibfield  {journal} {\bibinfo
  {journal} {Nature}\ }\textbf {\bibinfo {volume} {449}},\ \bibinfo {pages}
  {702} (\bibinfo {year} {2007})}\BibitemShut {NoStop}%
\bibitem [{\citenamefont {Spaldin}\ \emph {et~al.}(2008)\citenamefont
  {Spaldin}, \citenamefont {Fiebig},\ and\ \citenamefont
  {Mostovoy}}]{Spaldin2008-zj}%
  \BibitemOpen
  \bibfield  {author} {\bibinfo {author} {\bibfnamefont {N.~A.}\ \bibnamefont
  {Spaldin}}, \bibinfo {author} {\bibfnamefont {M.}~\bibnamefont {Fiebig}}, \
  and\ \bibinfo {author} {\bibfnamefont {M.}~\bibnamefont {Mostovoy}},\ }\href
  {https://iopscience.iop.org/article/10.1088/0953-8984/20/43/434203}
  {\bibfield  {journal} {\bibinfo  {journal} {J. Phys. Condens. Matter}\
  }\textbf {\bibinfo {volume} {20}},\ \bibinfo {pages} {434203} (\bibinfo
  {year} {2008})}\BibitemShut {NoStop}%
\bibitem [{\citenamefont {Zimmermann}\ \emph {et~al.}(2014)\citenamefont
  {Zimmermann}, \citenamefont {Meier},\ and\ \citenamefont
  {Fiebig}}]{Zimmermann2014-pv}%
  \BibitemOpen
  \bibfield  {author} {\bibinfo {author} {\bibfnamefont {A.~S.}\ \bibnamefont
  {Zimmermann}}, \bibinfo {author} {\bibfnamefont {D.}~\bibnamefont {Meier}}, \
  and\ \bibinfo {author} {\bibfnamefont {M.}~\bibnamefont {Fiebig}},\ }\href
  {http://dx.doi.org/10.1038/ncomms5796} {\bibfield  {journal} {\bibinfo
  {journal} {Nat. Commun.}\ }\textbf {\bibinfo {volume} {5}},\ \bibinfo {pages}
  {4796} (\bibinfo {year} {2014})}\BibitemShut {NoStop}%
\bibitem [{\citenamefont {Lim}\ \emph {et~al.}(2020)\citenamefont {Lim},
  \citenamefont {Varma}, \citenamefont {Eisaki},\ and\ \citenamefont
  {Kapitulnik}}]{Lim2020BSCCOphotocurrent}%
  \BibitemOpen
  \bibfield  {author} {\bibinfo {author} {\bibfnamefont {S.}~\bibnamefont
  {Lim}}, \bibinfo {author} {\bibfnamefont {C.~M.}\ \bibnamefont {Varma}},
  \bibinfo {author} {\bibfnamefont {H.}~\bibnamefont {Eisaki}}, \ and\ \bibinfo
  {author} {\bibfnamefont {A.}~\bibnamefont {Kapitulnik}},\ }\href
  {http://arxiv.org/abs/2011.06755} {\  (\bibinfo {year} {2020})},\ \Eprint
  {http://arxiv.org/abs/2011.06755} {arXiv:2011.06755 [cond-mat.str-el]}
  \BibitemShut {NoStop}%
\bibitem [{\citenamefont {Xu}\ \emph {et~al.}(2020{\natexlab{a}})\citenamefont
  {Xu}, \citenamefont {Ma}, \citenamefont {Gao}, \citenamefont {Kogar},
  \citenamefont {Zong}, \citenamefont {Mier~Valdivia}, \citenamefont {Dinh},
  \citenamefont {Huang}, \citenamefont {Singh}, \citenamefont {Hsu},
  \citenamefont {Chang}, \citenamefont {Ruff}, \citenamefont {Watanabe},
  \citenamefont {Taniguchi}, \citenamefont {Lin}, \citenamefont {Karapetrov},
  \citenamefont {Xiao}, \citenamefont {Jarillo-Herrero},\ and\ \citenamefont
  {Gedik}}]{Xu2020-ra}%
  \BibitemOpen
  \bibfield  {author} {\bibinfo {author} {\bibfnamefont {S.-Y.}\ \bibnamefont
  {Xu}}, \bibinfo {author} {\bibfnamefont {Q.}~\bibnamefont {Ma}}, \bibinfo
  {author} {\bibfnamefont {Y.}~\bibnamefont {Gao}}, \bibinfo {author}
  {\bibfnamefont {A.}~\bibnamefont {Kogar}}, \bibinfo {author} {\bibfnamefont
  {A.}~\bibnamefont {Zong}}, \bibinfo {author} {\bibfnamefont {A.~M.}\
  \bibnamefont {Mier~Valdivia}}, \bibinfo {author} {\bibfnamefont {T.~H.}\
  \bibnamefont {Dinh}}, \bibinfo {author} {\bibfnamefont {S.-M.}\ \bibnamefont
  {Huang}}, \bibinfo {author} {\bibfnamefont {B.}~\bibnamefont {Singh}},
  \bibinfo {author} {\bibfnamefont {C.-H.}\ \bibnamefont {Hsu}}, \bibinfo
  {author} {\bibfnamefont {T.-R.}\ \bibnamefont {Chang}}, \bibinfo {author}
  {\bibfnamefont {J.~P.~C.}\ \bibnamefont {Ruff}}, \bibinfo {author}
  {\bibfnamefont {K.}~\bibnamefont {Watanabe}}, \bibinfo {author}
  {\bibfnamefont {T.}~\bibnamefont {Taniguchi}}, \bibinfo {author}
  {\bibfnamefont {H.}~\bibnamefont {Lin}}, \bibinfo {author} {\bibfnamefont
  {G.}~\bibnamefont {Karapetrov}}, \bibinfo {author} {\bibfnamefont
  {D.}~\bibnamefont {Xiao}}, \bibinfo {author} {\bibfnamefont {P.}~\bibnamefont
  {Jarillo-Herrero}}, \ and\ \bibinfo {author} {\bibfnamefont {N.}~\bibnamefont
  {Gedik}},\ }\href {http://dx.doi.org/10.1038/s41586-020-2011-8} {\bibfield
  {journal} {\bibinfo  {journal} {Nature}\ }\textbf {\bibinfo {volume} {578}},\
  \bibinfo {pages} {545} (\bibinfo {year} {2020}{\natexlab{a}})}\BibitemShut
  {NoStop}%
\bibitem [{\citenamefont {Haldane}(1988)}]{Haldane1988-dm}%
  \BibitemOpen
  \bibfield  {author} {\bibinfo {author} {\bibfnamefont {F.~D.~M.}\
  \bibnamefont {Haldane}},\ }\href {\doibase 10.1103/PhysRevLett.61.2015}
  {\bibfield  {journal} {\bibinfo  {journal} {Phys. Rev. Lett.}\ }\textbf
  {\bibinfo {volume} {61}},\ \bibinfo {pages} {2015} (\bibinfo {year}
  {1988})}\BibitemShut {NoStop}%
\bibitem [{Note1()}]{Note1}%
  \BibitemOpen
  \bibinfo {note} {We note that the $\protect \mathcal {PT}$\protect \xspace
  symmetric current order in Fig.~\ref {Fig_honeycomb_orbital_order}(b) does
  not belong to the polar magnetic point group and cannot be denoted by the
  ferroic anapole order. It is labeled by the ordering of higher-rank magnetic
  multipole moments and toroidal multipole moments~\cite
  {Watanabe2018grouptheoretical,Hayami2018Classification}.}\BibitemShut {Stop}%
\bibitem [{\citenamefont {Izyumov}\ and\ \citenamefont
  {Ozerov}(1970)}]{izyumov2012magnetic}%
  \BibitemOpen
  \bibfield  {author} {\bibinfo {author} {\bibfnamefont {Y.~A.}\ \bibnamefont
  {Izyumov}}\ and\ \bibinfo {author} {\bibfnamefont {R.~P.}\ \bibnamefont
  {Ozerov}},\ }\href@noop {} {\emph {\bibinfo {title} {Magnetic neutron
  diffraction}}}\ (\bibinfo  {publisher} {Plenum},\ \bibinfo {address} {New
  York},\ \bibinfo {year} {1970})\BibitemShut {NoStop}%
\bibitem [{\citenamefont {Emery}(1987)}]{Emery1987}%
  \BibitemOpen
  \bibfield  {author} {\bibinfo {author} {\bibfnamefont {V.~J.}\ \bibnamefont
  {Emery}},\ }\href {\doibase 10.1103/PhysRevLett.58.2794} {\bibfield
  {journal} {\bibinfo  {journal} {Phys. Rev. Lett.}\ }\textbf {\bibinfo
  {volume} {58}},\ \bibinfo {pages} {2794} (\bibinfo {year}
  {1987})}\BibitemShut {NoStop}%
\bibitem [{\citenamefont {Ye}\ \emph {et~al.}(2015)\citenamefont {Ye},
  \citenamefont {Wang}, \citenamefont {Hoffmann}, \citenamefont {Wang},
  \citenamefont {Chi}, \citenamefont {Matsuda}, \citenamefont {Chakoumakos},
  \citenamefont {Fernandez-Baca},\ and\ \citenamefont {Cao}}]{Ye2015-vf}%
  \BibitemOpen
  \bibfield  {author} {\bibinfo {author} {\bibfnamefont {F.}~\bibnamefont
  {Ye}}, \bibinfo {author} {\bibfnamefont {X.}~\bibnamefont {Wang}}, \bibinfo
  {author} {\bibfnamefont {C.}~\bibnamefont {Hoffmann}}, \bibinfo {author}
  {\bibfnamefont {J.}~\bibnamefont {Wang}}, \bibinfo {author} {\bibfnamefont
  {S.}~\bibnamefont {Chi}}, \bibinfo {author} {\bibfnamefont {M.}~\bibnamefont
  {Matsuda}}, \bibinfo {author} {\bibfnamefont {B.~C.}\ \bibnamefont
  {Chakoumakos}}, \bibinfo {author} {\bibfnamefont {J.~A.}\ \bibnamefont
  {Fernandez-Baca}}, \ and\ \bibinfo {author} {\bibfnamefont {G.}~\bibnamefont
  {Cao}},\ }\href {\doibase 10.1103/PhysRevB.92.201112} {\bibfield  {journal}
  {\bibinfo  {journal} {Phys. Rev. B}\ }\textbf {\bibinfo {volume} {92}},\
  \bibinfo {pages} {201112} (\bibinfo {year} {2015})}\BibitemShut {NoStop}%
\bibitem [{\citenamefont {Crawford}\ \emph {et~al.}(1994)\citenamefont
  {Crawford}, \citenamefont {Subramanian}, \citenamefont {Harlow},
  \citenamefont {Fernandez-Baca}, \citenamefont {Wang},\ and\ \citenamefont
  {Johnston}}]{Crawford1994}%
  \BibitemOpen
  \bibfield  {author} {\bibinfo {author} {\bibfnamefont {M.~K.}\ \bibnamefont
  {Crawford}}, \bibinfo {author} {\bibfnamefont {M.~A.}\ \bibnamefont
  {Subramanian}}, \bibinfo {author} {\bibfnamefont {R.~L.}\ \bibnamefont
  {Harlow}}, \bibinfo {author} {\bibfnamefont {J.~A.}\ \bibnamefont
  {Fernandez-Baca}}, \bibinfo {author} {\bibfnamefont {Z.~R.}\ \bibnamefont
  {Wang}}, \ and\ \bibinfo {author} {\bibfnamefont {D.~C.}\ \bibnamefont
  {Johnston}},\ }\href {\doibase 10.1103/PhysRevB.49.9198} {\bibfield
  {journal} {\bibinfo  {journal} {Phys. Rev. B}\ }\textbf {\bibinfo {volume}
  {49}},\ \bibinfo {pages} {9198} (\bibinfo {year} {1994})}\BibitemShut
  {NoStop}%
\bibitem [{\citenamefont {Di~Matteo}\ and\ \citenamefont
  {Norman}(2016)}]{Matteo2016}%
  \BibitemOpen
  \bibfield  {author} {\bibinfo {author} {\bibfnamefont {S.}~\bibnamefont
  {Di~Matteo}}\ and\ \bibinfo {author} {\bibfnamefont {M.~R.}\ \bibnamefont
  {Norman}},\ }\href {\doibase 10.1103/PhysRevB.94.075148} {\bibfield
  {journal} {\bibinfo  {journal} {Phys. Rev. B}\ }\textbf {\bibinfo {volume}
  {94}},\ \bibinfo {pages} {075148} (\bibinfo {year} {2016})}\BibitemShut
  {NoStop}%
\bibitem [{\citenamefont {Kane}\ and\ \citenamefont {Mele}(2005)}]{Kane2005}%
  \BibitemOpen
  \bibfield  {author} {\bibinfo {author} {\bibfnamefont {C.~L.}\ \bibnamefont
  {Kane}}\ and\ \bibinfo {author} {\bibfnamefont {E.~J.}\ \bibnamefont
  {Mele}},\ }\href {\doibase 10.1103/PhysRevLett.95.226801} {\bibfield
  {journal} {\bibinfo  {journal} {Phys. Rev. Lett.}\ }\textbf {\bibinfo
  {volume} {95}},\ \bibinfo {pages} {226801} (\bibinfo {year}
  {2005})}\BibitemShut {NoStop}%
\bibitem [{\citenamefont {Kim}\ \emph {et~al.}(2009)\citenamefont {Kim},
  \citenamefont {Ohsumi}, \citenamefont {Komesu}, \citenamefont {Sakai},
  \citenamefont {Morita}, \citenamefont {Takagi},\ and\ \citenamefont
  {Arima}}]{Kim2009}%
  \BibitemOpen
  \bibfield  {author} {\bibinfo {author} {\bibfnamefont {B.~J.}\ \bibnamefont
  {Kim}}, \bibinfo {author} {\bibfnamefont {H.}~\bibnamefont {Ohsumi}},
  \bibinfo {author} {\bibfnamefont {T.}~\bibnamefont {Komesu}}, \bibinfo
  {author} {\bibfnamefont {S.}~\bibnamefont {Sakai}}, \bibinfo {author}
  {\bibfnamefont {T.}~\bibnamefont {Morita}}, \bibinfo {author} {\bibfnamefont
  {H.}~\bibnamefont {Takagi}}, \ and\ \bibinfo {author} {\bibfnamefont
  {T.}~\bibnamefont {Arima}},\ }\href
  {http://dx.doi.org/10.1126/science.1167106} {\bibfield  {journal} {\bibinfo
  {journal} {Science}\ }\textbf {\bibinfo {volume} {323}},\ \bibinfo {pages}
  {1329} (\bibinfo {year} {2009})}\BibitemShut {NoStop}%
\bibitem [{\citenamefont {Dhital}\ \emph {et~al.}(2013)\citenamefont {Dhital},
  \citenamefont {Hogan}, \citenamefont {Yamani}, \citenamefont {de~la Cruz},
  \citenamefont {Chen}, \citenamefont {Khadka}, \citenamefont {Ren},\ and\
  \citenamefont {Wilson}}]{Dhital2013}%
  \BibitemOpen
  \bibfield  {author} {\bibinfo {author} {\bibfnamefont {C.}~\bibnamefont
  {Dhital}}, \bibinfo {author} {\bibfnamefont {T.}~\bibnamefont {Hogan}},
  \bibinfo {author} {\bibfnamefont {Z.}~\bibnamefont {Yamani}}, \bibinfo
  {author} {\bibfnamefont {C.}~\bibnamefont {de~la Cruz}}, \bibinfo {author}
  {\bibfnamefont {X.}~\bibnamefont {Chen}}, \bibinfo {author} {\bibfnamefont
  {S.}~\bibnamefont {Khadka}}, \bibinfo {author} {\bibfnamefont
  {Z.}~\bibnamefont {Ren}}, \ and\ \bibinfo {author} {\bibfnamefont {S.~D.}\
  \bibnamefont {Wilson}},\ }\href {\doibase 10.1103/PhysRevB.87.144405}
  {\bibfield  {journal} {\bibinfo  {journal} {Phys. Rev. B}\ }\textbf {\bibinfo
  {volume} {87}},\ \bibinfo {pages} {144405} (\bibinfo {year}
  {2013})}\BibitemShut {NoStop}%
\bibitem [{\citenamefont {Shitade}\ \emph {et~al.}(2018)\citenamefont
  {Shitade}, \citenamefont {Watanabe},\ and\ \citenamefont
  {Yanase}}]{Shitade2018}%
  \BibitemOpen
  \bibfield  {author} {\bibinfo {author} {\bibfnamefont {A.}~\bibnamefont
  {Shitade}}, \bibinfo {author} {\bibfnamefont {H.}~\bibnamefont {Watanabe}}, \
  and\ \bibinfo {author} {\bibfnamefont {Y.}~\bibnamefont {Yanase}},\ }\href
  {\doibase 10.1103/PhysRevB.98.020407} {\bibfield  {journal} {\bibinfo
  {journal} {Phys. Rev. B}\ }\textbf {\bibinfo {volume} {98}},\ \bibinfo
  {pages} {020407} (\bibinfo {year} {2018})}\BibitemShut {NoStop}%
\bibitem [{\citenamefont {Gao}\ and\ \citenamefont
  {Xiao}(2018)}]{Gao2018orbitalMQM}%
  \BibitemOpen
  \bibfield  {author} {\bibinfo {author} {\bibfnamefont {Y.}~\bibnamefont
  {Gao}}\ and\ \bibinfo {author} {\bibfnamefont {D.}~\bibnamefont {Xiao}},\
  }\href {\doibase 10.1103/PhysRevB.98.060402} {\bibfield  {journal} {\bibinfo
  {journal} {Phys. Rev. B}\ }\textbf {\bibinfo {volume} {98}},\ \bibinfo
  {pages} {060402} (\bibinfo {year} {2018})}\BibitemShut {NoStop}%
\bibitem [{\citenamefont {Watanabe}\ \emph {et~al.}(2010)\citenamefont
  {Watanabe}, \citenamefont {Shirakawa},\ and\ \citenamefont
  {Yunoki}}]{HiroshiWatanabe2010}%
  \BibitemOpen
  \bibfield  {author} {\bibinfo {author} {\bibfnamefont {H.}~\bibnamefont
  {Watanabe}}, \bibinfo {author} {\bibfnamefont {T.}~\bibnamefont {Shirakawa}},
  \ and\ \bibinfo {author} {\bibfnamefont {S.}~\bibnamefont {Yunoki}},\ }\href
  {\doibase 10.1103/PhysRevLett.105.216410} {\bibfield  {journal} {\bibinfo
  {journal} {Phys. Rev. Lett.}\ }\textbf {\bibinfo {volume} {105}},\ \bibinfo
  {pages} {216410} (\bibinfo {year} {2010})}\BibitemShut {NoStop}%
\bibitem [{\citenamefont {Kim}\ \emph {et~al.}(2008)\citenamefont {Kim},
  \citenamefont {Jin}, \citenamefont {Moon}, \citenamefont {Kim}, \citenamefont
  {Park}, \citenamefont {Leem}, \citenamefont {Yu}, \citenamefont {Noh},
  \citenamefont {Kim}, \citenamefont {Oh}, \citenamefont {Park}, \citenamefont
  {Durairaj}, \citenamefont {Cao},\ and\ \citenamefont
  {Rotenberg}}]{Kim2008SOCMott}%
  \BibitemOpen
  \bibfield  {author} {\bibinfo {author} {\bibfnamefont {B.~J.}\ \bibnamefont
  {Kim}}, \bibinfo {author} {\bibfnamefont {H.}~\bibnamefont {Jin}}, \bibinfo
  {author} {\bibfnamefont {S.~J.}\ \bibnamefont {Moon}}, \bibinfo {author}
  {\bibfnamefont {J.-Y.}\ \bibnamefont {Kim}}, \bibinfo {author} {\bibfnamefont
  {B.-G.}\ \bibnamefont {Park}}, \bibinfo {author} {\bibfnamefont {C.~S.}\
  \bibnamefont {Leem}}, \bibinfo {author} {\bibfnamefont {J.}~\bibnamefont
  {Yu}}, \bibinfo {author} {\bibfnamefont {T.~W.}\ \bibnamefont {Noh}},
  \bibinfo {author} {\bibfnamefont {C.}~\bibnamefont {Kim}}, \bibinfo {author}
  {\bibfnamefont {S.-J.}\ \bibnamefont {Oh}}, \bibinfo {author} {\bibfnamefont
  {J.-H.}\ \bibnamefont {Park}}, \bibinfo {author} {\bibfnamefont
  {V.}~\bibnamefont {Durairaj}}, \bibinfo {author} {\bibfnamefont
  {G.}~\bibnamefont {Cao}}, \ and\ \bibinfo {author} {\bibfnamefont
  {E.}~\bibnamefont {Rotenberg}},\ }\href {\doibase
  10.1103/PhysRevLett.101.076402} {\bibfield  {journal} {\bibinfo  {journal}
  {Phys. Rev. Lett.}\ }\textbf {\bibinfo {volume} {101}},\ \bibinfo {pages}
  {076402} (\bibinfo {year} {2008})}\BibitemShut {NoStop}%
\bibitem [{\citenamefont {Souza}\ and\ \citenamefont
  {Vanderbilt}(2008)}]{Souza2008}%
  \BibitemOpen
  \bibfield  {author} {\bibinfo {author} {\bibfnamefont {I.}~\bibnamefont
  {Souza}}\ and\ \bibinfo {author} {\bibfnamefont {D.}~\bibnamefont
  {Vanderbilt}},\ }\href {\doibase 10.1103/PhysRevB.77.054438} {\bibfield
  {journal} {\bibinfo  {journal} {Phys. Rev. B}\ }\textbf {\bibinfo {volume}
  {77}},\ \bibinfo {pages} {054438} (\bibinfo {year} {2008})}\BibitemShut
  {NoStop}%
\bibitem [{\citenamefont {von Baltz}\ and\ \citenamefont
  {Kraut}(1981)}]{Kraut1981Photovoltaiceffect}%
  \BibitemOpen
  \bibfield  {author} {\bibinfo {author} {\bibfnamefont {R.}~\bibnamefont {von
  Baltz}}\ and\ \bibinfo {author} {\bibfnamefont {W.}~\bibnamefont {Kraut}},\
  }\href {\doibase 10.1103/PhysRevB.23.5590} {\bibfield  {journal} {\bibinfo
  {journal} {Phys. Rev. B}\ }\textbf {\bibinfo {volume} {23}},\ \bibinfo
  {pages} {5590} (\bibinfo {year} {1981})}\BibitemShut {NoStop}%
\bibitem [{\citenamefont {Grosso}\ and\ \citenamefont
  {Parravicini}(2013)}]{Grosso2013Book}%
  \BibitemOpen
  \bibfield  {author} {\bibinfo {author} {\bibfnamefont {G.}~\bibnamefont
  {Grosso}}\ and\ \bibinfo {author} {\bibfnamefont {G.}~\bibnamefont
  {Parravicini}},\ }\href@noop {} {\emph {\bibinfo {title} {Solid State
  Physics, 2nd Edition}}}\ (\bibinfo  {publisher} {Academic Press},\ \bibinfo
  {address} {Amsterdam},\ \bibinfo {year} {2013})\BibitemShut {NoStop}%
\bibitem [{\citenamefont {Hatada}\ \emph {et~al.}(2020)\citenamefont {Hatada},
  \citenamefont {Nakamura}, \citenamefont {Sotome}, \citenamefont {Kaneko},
  \citenamefont {Ogawa}, \citenamefont {Morimoto}, \citenamefont {Tokura},\
  and\ \citenamefont {Kawasaki}}]{Hatada2020}%
  \BibitemOpen
  \bibfield  {author} {\bibinfo {author} {\bibfnamefont {H.}~\bibnamefont
  {Hatada}}, \bibinfo {author} {\bibfnamefont {M.}~\bibnamefont {Nakamura}},
  \bibinfo {author} {\bibfnamefont {M.}~\bibnamefont {Sotome}}, \bibinfo
  {author} {\bibfnamefont {Y.}~\bibnamefont {Kaneko}}, \bibinfo {author}
  {\bibfnamefont {N.}~\bibnamefont {Ogawa}}, \bibinfo {author} {\bibfnamefont
  {T.}~\bibnamefont {Morimoto}}, \bibinfo {author} {\bibfnamefont
  {Y.}~\bibnamefont {Tokura}}, \ and\ \bibinfo {author} {\bibfnamefont
  {M.}~\bibnamefont {Kawasaki}},\ }\href
  {http://dx.doi.org/10.1073/pnas.2007002117} {\bibfield  {journal} {\bibinfo
  {journal} {Proc. Natl. Acad. Sci. U. S. A.}\ }\textbf {\bibinfo {volume}
  {117}},\ \bibinfo {pages} {20411} (\bibinfo {year} {2020})}\BibitemShut
  {NoStop}%
\bibitem [{\citenamefont {Nastos}\ and\ \citenamefont
  {Sipe}(2006)}]{Nastos2006}%
  \BibitemOpen
  \bibfield  {author} {\bibinfo {author} {\bibfnamefont {F.}~\bibnamefont
  {Nastos}}\ and\ \bibinfo {author} {\bibfnamefont {J.~E.}\ \bibnamefont
  {Sipe}},\ }\href {\doibase 10.1103/PhysRevB.74.035201} {\bibfield  {journal}
  {\bibinfo  {journal} {Phys. Rev. B}\ }\textbf {\bibinfo {volume} {74}},\
  \bibinfo {pages} {035201} (\bibinfo {year} {2006})}\BibitemShut {NoStop}%
\bibitem [{\citenamefont {Iba\~nez Azpiroz}\ \emph {et~al.}(2018)\citenamefont
  {Iba\~nez Azpiroz}, \citenamefont {Tsirkin},\ and\ \citenamefont
  {Souza}}]{Ibanez2018}%
  \BibitemOpen
  \bibfield  {author} {\bibinfo {author} {\bibfnamefont {J.}~\bibnamefont
  {Iba\~nez Azpiroz}}, \bibinfo {author} {\bibfnamefont {S.~S.}\ \bibnamefont
  {Tsirkin}}, \ and\ \bibinfo {author} {\bibfnamefont {I.}~\bibnamefont
  {Souza}},\ }\href {\doibase 10.1103/PhysRevB.97.245143} {\bibfield  {journal}
  {\bibinfo  {journal} {Phys. Rev. B}\ }\textbf {\bibinfo {volume} {97}},\
  \bibinfo {pages} {245143} (\bibinfo {year} {2018})}\BibitemShut {NoStop}%
\bibitem [{\citenamefont {Kastl}\ \emph {et~al.}(2015)\citenamefont {Kastl},
  \citenamefont {Karnetzky}, \citenamefont {Karl},\ and\ \citenamefont
  {Holleitner}}]{Kastl2015-rj}%
  \BibitemOpen
  \bibfield  {author} {\bibinfo {author} {\bibfnamefont {C.}~\bibnamefont
  {Kastl}}, \bibinfo {author} {\bibfnamefont {C.}~\bibnamefont {Karnetzky}},
  \bibinfo {author} {\bibfnamefont {H.}~\bibnamefont {Karl}}, \ and\ \bibinfo
  {author} {\bibfnamefont {A.~W.}\ \bibnamefont {Holleitner}},\ }\href
  {http://dx.doi.org/10.1038/ncomms7617} {\bibfield  {journal} {\bibinfo
  {journal} {Nat. Commun.}\ }\textbf {\bibinfo {volume} {6}},\ \bibinfo {pages}
  {6617} (\bibinfo {year} {2015})}\BibitemShut {NoStop}%
\bibitem [{\citenamefont {Takeno}\ \emph {et~al.}(2018)\citenamefont {Takeno},
  \citenamefont {Saito},\ and\ \citenamefont {Mizoguchi}}]{Takeno2018-gy}%
  \BibitemOpen
  \bibfield  {author} {\bibinfo {author} {\bibfnamefont {H.}~\bibnamefont
  {Takeno}}, \bibinfo {author} {\bibfnamefont {S.}~\bibnamefont {Saito}}, \
  and\ \bibinfo {author} {\bibfnamefont {K.}~\bibnamefont {Mizoguchi}},\ }\href
  {http://dx.doi.org/10.1038/s41598-018-33716-0} {\bibfield  {journal}
  {\bibinfo  {journal} {Sci. Rep.}\ }\textbf {\bibinfo {volume} {8}},\ \bibinfo
  {pages} {15392} (\bibinfo {year} {2018})}\BibitemShut {NoStop}%
\bibitem [{\citenamefont {Sotome}\ \emph {et~al.}(2019)\citenamefont {Sotome},
  \citenamefont {Nakamura}, \citenamefont {Fujioka}, \citenamefont {Ogino},
  \citenamefont {Kaneko}, \citenamefont {Morimoto}, \citenamefont {Zhang},
  \citenamefont {Kawasaki}, \citenamefont {Nagaosa}, \citenamefont {Tokura},\
  and\ \citenamefont {Ogawa}}]{Sotome2019}%
  \BibitemOpen
  \bibfield  {author} {\bibinfo {author} {\bibfnamefont {M.}~\bibnamefont
  {Sotome}}, \bibinfo {author} {\bibfnamefont {M.}~\bibnamefont {Nakamura}},
  \bibinfo {author} {\bibfnamefont {J.}~\bibnamefont {Fujioka}}, \bibinfo
  {author} {\bibfnamefont {M.}~\bibnamefont {Ogino}}, \bibinfo {author}
  {\bibfnamefont {Y.}~\bibnamefont {Kaneko}}, \bibinfo {author} {\bibfnamefont
  {T.}~\bibnamefont {Morimoto}}, \bibinfo {author} {\bibfnamefont
  {Y.}~\bibnamefont {Zhang}}, \bibinfo {author} {\bibfnamefont
  {M.}~\bibnamefont {Kawasaki}}, \bibinfo {author} {\bibfnamefont
  {N.}~\bibnamefont {Nagaosa}}, \bibinfo {author} {\bibfnamefont
  {Y.}~\bibnamefont {Tokura}}, \ and\ \bibinfo {author} {\bibfnamefont
  {N.}~\bibnamefont {Ogawa}},\ }\href {https://doi.org/10.1063/1.5087960}
  {\bibfield  {journal} {\bibinfo  {journal} {Appl. Phys. Lett.}\ }\textbf
  {\bibinfo {volume} {114}},\ \bibinfo {pages} {151101} (\bibinfo {year}
  {2019})}\BibitemShut {NoStop}%
\bibitem [{\citenamefont {Sirica}\ \emph {et~al.}(2019)\citenamefont {Sirica},
  \citenamefont {Tobey}, \citenamefont {Zhao}, \citenamefont {Chen},
  \citenamefont {Xu}, \citenamefont {Yang}, \citenamefont {Shen}, \citenamefont
  {Yarotski}, \citenamefont {Bowlan}, \citenamefont {Trugman}, \citenamefont
  {Zhu}, \citenamefont {Dai}, \citenamefont {Azad}, \citenamefont {Ni},
  \citenamefont {Qiu}, \citenamefont {Taylor},\ and\ \citenamefont
  {Prasankumar}}]{Sirica2019Ultrafast}%
  \BibitemOpen
  \bibfield  {author} {\bibinfo {author} {\bibfnamefont {N.}~\bibnamefont
  {Sirica}}, \bibinfo {author} {\bibfnamefont {R.~I.}\ \bibnamefont {Tobey}},
  \bibinfo {author} {\bibfnamefont {L.~X.}\ \bibnamefont {Zhao}}, \bibinfo
  {author} {\bibfnamefont {G.~F.}\ \bibnamefont {Chen}}, \bibinfo {author}
  {\bibfnamefont {B.}~\bibnamefont {Xu}}, \bibinfo {author} {\bibfnamefont
  {R.}~\bibnamefont {Yang}}, \bibinfo {author} {\bibfnamefont {B.}~\bibnamefont
  {Shen}}, \bibinfo {author} {\bibfnamefont {D.~A.}\ \bibnamefont {Yarotski}},
  \bibinfo {author} {\bibfnamefont {P.}~\bibnamefont {Bowlan}}, \bibinfo
  {author} {\bibfnamefont {S.~A.}\ \bibnamefont {Trugman}}, \bibinfo {author}
  {\bibfnamefont {J.-X.}\ \bibnamefont {Zhu}}, \bibinfo {author} {\bibfnamefont
  {Y.~M.}\ \bibnamefont {Dai}}, \bibinfo {author} {\bibfnamefont {A.~K.}\
  \bibnamefont {Azad}}, \bibinfo {author} {\bibfnamefont {N.}~\bibnamefont
  {Ni}}, \bibinfo {author} {\bibfnamefont {X.~G.}\ \bibnamefont {Qiu}},
  \bibinfo {author} {\bibfnamefont {A.~J.}\ \bibnamefont {Taylor}}, \ and\
  \bibinfo {author} {\bibfnamefont {R.~P.}\ \bibnamefont {Prasankumar}},\
  }\href {\doibase 10.1103/PhysRevLett.122.197401} {\bibfield  {journal}
  {\bibinfo  {journal} {Phys. Rev. Lett.}\ }\textbf {\bibinfo {volume} {122}},\
  \bibinfo {pages} {197401} (\bibinfo {year} {2019})}\BibitemShut {NoStop}%
\bibitem [{\citenamefont {Gao}\ \emph {et~al.}(2020)\citenamefont {Gao},
  \citenamefont {Kaushik}, \citenamefont {Philip}, \citenamefont {Li},
  \citenamefont {Qin}, \citenamefont {Liu}, \citenamefont {Zhang},
  \citenamefont {Su}, \citenamefont {Chen}, \citenamefont {Weng}, \citenamefont
  {Kharzeev}, \citenamefont {Liu},\ and\ \citenamefont {Qi}}]{Gao2020-du}%
  \BibitemOpen
  \bibfield  {author} {\bibinfo {author} {\bibfnamefont {Y.}~\bibnamefont
  {Gao}}, \bibinfo {author} {\bibfnamefont {S.}~\bibnamefont {Kaushik}},
  \bibinfo {author} {\bibfnamefont {E.~J.}\ \bibnamefont {Philip}}, \bibinfo
  {author} {\bibfnamefont {Z.}~\bibnamefont {Li}}, \bibinfo {author}
  {\bibfnamefont {Y.}~\bibnamefont {Qin}}, \bibinfo {author} {\bibfnamefont
  {Y.~P.}\ \bibnamefont {Liu}}, \bibinfo {author} {\bibfnamefont {W.~L.}\
  \bibnamefont {Zhang}}, \bibinfo {author} {\bibfnamefont {Y.~L.}\ \bibnamefont
  {Su}}, \bibinfo {author} {\bibfnamefont {X.}~\bibnamefont {Chen}}, \bibinfo
  {author} {\bibfnamefont {H.}~\bibnamefont {Weng}}, \bibinfo {author}
  {\bibfnamefont {D.~E.}\ \bibnamefont {Kharzeev}}, \bibinfo {author}
  {\bibfnamefont {M.~K.}\ \bibnamefont {Liu}}, \ and\ \bibinfo {author}
  {\bibfnamefont {J.}~\bibnamefont {Qi}},\ }\href
  {http://dx.doi.org/10.1038/s41467-020-14463-1} {\bibfield  {journal}
  {\bibinfo  {journal} {Nat. Commun.}\ }\textbf {\bibinfo {volume} {11}},\
  \bibinfo {pages} {720} (\bibinfo {year} {2020})}\BibitemShut {NoStop}%
\bibitem [{\citenamefont {Xu}\ \emph {et~al.}(2020{\natexlab{b}})\citenamefont
  {Xu}, \citenamefont {Marsik}, \citenamefont {Sheveleva}, \citenamefont
  {Lyzwa}, \citenamefont {Louat}, \citenamefont {Brouet}, \citenamefont
  {Munzar},\ and\ \citenamefont {Bernhard}}]{Xu2020}%
  \BibitemOpen
  \bibfield  {author} {\bibinfo {author} {\bibfnamefont {B.}~\bibnamefont
  {Xu}}, \bibinfo {author} {\bibfnamefont {P.}~\bibnamefont {Marsik}}, \bibinfo
  {author} {\bibfnamefont {E.}~\bibnamefont {Sheveleva}}, \bibinfo {author}
  {\bibfnamefont {F.}~\bibnamefont {Lyzwa}}, \bibinfo {author} {\bibfnamefont
  {A.}~\bibnamefont {Louat}}, \bibinfo {author} {\bibfnamefont
  {V.}~\bibnamefont {Brouet}}, \bibinfo {author} {\bibfnamefont
  {D.}~\bibnamefont {Munzar}}, \ and\ \bibinfo {author} {\bibfnamefont
  {C.}~\bibnamefont {Bernhard}},\ }\href
  {http://dx.doi.org/10.1103/PhysRevLett.124.027402} {\bibfield  {journal}
  {\bibinfo  {journal} {Phys. Rev. Lett.}\ }\textbf {\bibinfo {volume} {124}},\
  \bibinfo {pages} {027402} (\bibinfo {year} {2020}{\natexlab{b}})}\BibitemShut
  {NoStop}%
\bibitem [{\citenamefont {Cao}\ \emph {et~al.}(2016)\citenamefont {Cao},
  \citenamefont {Wang}, \citenamefont {Waugh}, \citenamefont {Reber},
  \citenamefont {Li}, \citenamefont {Zhou}, \citenamefont {Parham},
  \citenamefont {Park}, \citenamefont {Plumb}, \citenamefont {Rotenberg},
  \citenamefont {Bostwick}, \citenamefont {Denlinger}, \citenamefont {Qi},
  \citenamefont {Hermele}, \citenamefont {Cao},\ and\ \citenamefont
  {Dessau}}]{Cao2016}%
  \BibitemOpen
  \bibfield  {author} {\bibinfo {author} {\bibfnamefont {Y.}~\bibnamefont
  {Cao}}, \bibinfo {author} {\bibfnamefont {Q.}~\bibnamefont {Wang}}, \bibinfo
  {author} {\bibfnamefont {J.~A.}\ \bibnamefont {Waugh}}, \bibinfo {author}
  {\bibfnamefont {T.~J.}\ \bibnamefont {Reber}}, \bibinfo {author}
  {\bibfnamefont {H.}~\bibnamefont {Li}}, \bibinfo {author} {\bibfnamefont
  {X.}~\bibnamefont {Zhou}}, \bibinfo {author} {\bibfnamefont {S.}~\bibnamefont
  {Parham}}, \bibinfo {author} {\bibfnamefont {S.-R.}\ \bibnamefont {Park}},
  \bibinfo {author} {\bibfnamefont {N.~C.}\ \bibnamefont {Plumb}}, \bibinfo
  {author} {\bibfnamefont {E.}~\bibnamefont {Rotenberg}}, \bibinfo {author}
  {\bibfnamefont {A.}~\bibnamefont {Bostwick}}, \bibinfo {author}
  {\bibfnamefont {J.~D.}\ \bibnamefont {Denlinger}}, \bibinfo {author}
  {\bibfnamefont {T.}~\bibnamefont {Qi}}, \bibinfo {author} {\bibfnamefont
  {M.~A.}\ \bibnamefont {Hermele}}, \bibinfo {author} {\bibfnamefont
  {G.}~\bibnamefont {Cao}}, \ and\ \bibinfo {author} {\bibfnamefont {D.~S.}\
  \bibnamefont {Dessau}},\ }\href {http://dx.doi.org/10.1038/ncomms11367}
  {\bibfield  {journal} {\bibinfo  {journal} {Nat. Commun.}\ }\textbf {\bibinfo
  {volume} {7}},\ \bibinfo {pages} {11367} (\bibinfo {year}
  {2016})}\BibitemShut {NoStop}%
\bibitem [{\citenamefont {Louat}\ \emph {et~al.}(2018)\citenamefont {Louat},
  \citenamefont {Bert}, \citenamefont {Serrier-Garcia}, \citenamefont
  {Bertran}, \citenamefont {Le~F\`evre}, \citenamefont {Rault},\ and\
  \citenamefont {Brouet}}]{Louat2018}%
  \BibitemOpen
  \bibfield  {author} {\bibinfo {author} {\bibfnamefont {A.}~\bibnamefont
  {Louat}}, \bibinfo {author} {\bibfnamefont {F.}~\bibnamefont {Bert}},
  \bibinfo {author} {\bibfnamefont {L.}~\bibnamefont {Serrier-Garcia}},
  \bibinfo {author} {\bibfnamefont {F.}~\bibnamefont {Bertran}}, \bibinfo
  {author} {\bibfnamefont {P.}~\bibnamefont {Le~F\`evre}}, \bibinfo {author}
  {\bibfnamefont {J.}~\bibnamefont {Rault}}, \ and\ \bibinfo {author}
  {\bibfnamefont {V.}~\bibnamefont {Brouet}},\ }\href {\doibase
  10.1103/PhysRevB.97.161109} {\bibfield  {journal} {\bibinfo  {journal} {Phys.
  Rev. B}\ }\textbf {\bibinfo {volume} {97}},\ \bibinfo {pages} {161109}
  (\bibinfo {year} {2018})}\BibitemShut {NoStop}%
\bibitem [{\citenamefont {Mitchell}(2015)}]{Mitchell2015}%
  \BibitemOpen
  \bibfield  {author} {\bibinfo {author} {\bibfnamefont {J.~F.}\ \bibnamefont
  {Mitchell}},\ }\href {https://doi.org/10.1063/1.4921953} {\bibfield
  {journal} {\bibinfo  {journal} {APL Materials}\ }\textbf {\bibinfo {volume}
  {3}},\ \bibinfo {pages} {062404} (\bibinfo {year} {2015})}\BibitemShut
  {NoStop}%
\end{thebibliography}
\end{document}